\documentclass[conference]{IEEEtran}
\IEEEoverridecommandlockouts

\usepackage{cite}
\usepackage{amsmath,amssymb}
\usepackage{booktabs}
\usepackage{graphicx}
\usepackage{xcolor}
\usepackage{url}
\usepackage{microtype}
\usepackage[T1]{fontenc}
\usepackage[utf8]{inputenc}
\usepackage{multirow}
\usepackage[hidelinks]{hyperref}
\hypersetup{
  pdftitle={CTTE: An Open Dual-Protocol RISC-V Trace Encoder for N-Trace and E-Trace},
  pdfauthor={Alexander Weiss and Albert Schulz},
  pdfsubject={Open RISC-V processor trace encoder with N-Trace and E-Trace back ends},
  pdfkeywords={RISC-V, N-Trace, E-Trace, processor trace, open hardware}
}

\newif\ifdraftmode

\ifdraftmode
  
  \newcommand{\note}[1]{\textcolor{blue}{\small\textsf{[#1]}}}
\else
  
  \newcommand{\note}[1]{}
\fi

\newcommand{\ctte}{CTTE}

\newcommand{\bpi}{\ensuremath{\mathrm{bpi}}}

\begin{document}

\title{\ctte: An Open Dual-Protocol RISC-V Trace Encoder for N-Trace and E-Trace}

\author{\IEEEauthorblockN{Alexander Weiss and Albert Schulz}
\IEEEauthorblockA{\textit{Accemic Technologies GmbH} \\
Kiefersfelden, Germany \\
\{aweiss, aschulz\}@accemic.com}}

\maketitle

\begin{abstract}
RISC-V defines two ratified processor-trace formats, N-Trace and E-Trace, that
share a hart-to-encoder instruction trace interface but differ in compression,
message representation, and framing. To the best of our knowledge, as of August
2026, no publicly available synthesizable N-Trace encoder and no published
hardware encoder with both back ends from a common front end have previously
been reported. We present \ctte{} (CEDARtools.TraceEncoder), an open
SystemVerilog trace encoder with a protocol-agnostic front end and selectable
N-Trace/Nexus or E-Trace back ends. \ctte{} implements the N-Trace 1.0
program-trace message set in both instruction-trace modes, supports
parameterized N-Trace address width, follows the RISC-V Trace Control
Interface, and treats output-bandwidth loss as an explicitly announced event
followed by state re-convergence on the evaluated N-Trace path. \ctte{} has been integrated with six RISC-V cores from five suppliers. We
evaluate it here on 64-bit RISC-V systems booting Linux, including a two-hart
SMP configuration, and demonstrate
source-side process-context filtering that doubles the observation depth for a
target process in a fixed capture buffer. Verification combines
instruction-exact round trips, standing RTL invariants, formal model checking,
and machine-judged hardware campaigns with negative controls. With hardware and
encoder configuration held constant, workload choice alone changes measured
trace cost from 0.21 to 4.90 bits per retired instruction. A common front end
additionally enables controlled N-Trace/E-Trace back-end comparisons while
exposing the residual synchronization and transport effects that must be
separated from the wire format itself. RTL, register sources, testbenches,
formal properties, decoder extensions, and build scripts are released under
CERN-OHL-S-2.0; proprietary integration is available under an Accemic commercial
license.
\end{abstract}


\section{Introduction}

Instruction trace is a principal mechanism for reporting what a processor
\emph{actually did}, continuously, at instruction granularity, without
instrumenting the observed program. Everything built on top of
it---interactive debug of non-reproducible failures, hybrid worst-case-execution-time
estimation, structural-coverage evidence for a certification argument, runtime
verification of a deployed system---inherits the trustworthiness of the encoder
that produced the stream. This gives the encoder an unusual status among
IP blocks: a debugger that crashes costs an afternoon, but a trace instrument
that silently drops or reorders one message invalidates every conclusion drawn
from the capture, and does so without any signal that it happened.

RISC-V has standardized this path twice. The \emph{hart-to-encoder} side is
shared: both specifications describe the same instruction trace interface
(ITI, historically also called the trace ingress port or TIP), carrying
per-retirement \texttt{itype}, \texttt{iaddr}, \texttt{iretire},
\texttt{ilastsize}, privilege and context information. The \emph{encoder-to-sink}
side is not. \emph{Efficient Trace for RISC-V} (E-Trace)~\cite{riscv-etrace}
defines its own \texttt{te\_inst} payloads and encapsulation.
\emph{N-Trace}~\cite{riscv-ntrace}, ratified in November 2024, keeps the same
ingress but emits Nexus messages per IEEE-ISTO 5001-2012~\cite{nexus5001}:
TCODE-identified messages framed by the self-describing MDO/MSEO encoding.
A third specification, the Trace Control Interface (TCI)~\cite{riscv-tci},
defines the register-level programming model for either.

The published implementation record is thin, and thinner still if one asks for
artifacts rather than results. Kükner et al.~\cite{kukner2022} demonstrated
multi-retirement encoding in 28\,nm; a Chalmers thesis~\cite{hessman2024}
implemented E-Trace in VHDL for NOEL-V and verified it against a third-party
decoder; Laghi et al.~\cite{laghi2025} integrated an E-Trace tracing system into
a CVA6-based platform and subsequently released the RTL~\cite{rvtracer};
Janarthanam et al.~\cite{janarthanam2025} described an N-Trace encoder for a
high-performance out-of-order core with a two-oracle verification methodology;
and Karslioglu and Akturk~\cite{karslioglu2025} designed and evaluated an
N-Trace tracer. To the best of our knowledge (search date 2026-08-09), none of
the N-Trace hardware implementations is available as public synthesizable
source, and no published design produces \emph{both} standardized formats.

A second gap is in what gets evaluated. Published encoder evaluations reviewed
in our search as of August 2026 trace bare-metal or single-application
workloads on a single hart; we found no evaluation of the regime that actually
stresses a trace subsystem: a virtual-memory
operating system on several harts, where the address space is 64-bit and
sparse, where the process under observation is descheduled and migrated between
cores, and where the interesting execution is a small fraction of everything the
trace port carries. That regime breaks assumptions on both sides of the link. On
the encoder side it decides whether an architectural process identifier is even
available and stable across harts. On the decoder side we found that it breaks
tools outright: with the unmodified reference decoder, our Linux capture
terminated at the first return-stack mismatch after a scheduler
switch---precisely the event a multi-process trace exists to show.

That last gap is more than a feature checkbox. Because the two formats consume
the identical ingress, a single design that drives both back ends from one
shared front end is a measurement instrument: it isolates the cost that is
attributable to the \emph{wire format} from the cost attributable to the
front-end architecture, on the same retirement stream, in the same netlist,
with the same synthesis flow. Cross-paper comparisons cannot do this, because
compression rate depends on the synchronization cadence, the enabled optional
features, the treatment of framing bytes and the workload mix at least as
strongly as on the format itself.

\subsection*{Research question}

\emph{What does it cost---in area, in timing, in bandwidth and in verification
effort---to build a RISC-V trace encoder that stays decodable under the
conditions a real system imposes: 64-bit sparse address spaces, several harts,
a preemptive operating system, and an output port that stalls? And how do
N-Trace and E-Trace differ when driven by the same recorded ingress stream and
the same front-end configuration, and which residual back-end effects limit
attribution of the measured difference to the wire format itself?}

\subsection*{Contributions}

This work makes five contributions. \textbf{First}, \ctte{} is an open,
synthesizable trace encoder whose protocol-agnostic front end drives either an
N-Trace/Nexus or an E-Trace back end, making the back-end choice independently
measurable. \textbf{Second}, we evaluate trace in the 64-bit, multi-hart
operating-system regime and demonstrate source-side process-context filtering on
hardware. \textbf{Third}, we define an announced-loss and re-convergence
contract and support it with instruction-exact round trips, standing RTL
invariants, formal model checking, and machine-judged hardware campaigns with
negative controls. \textbf{Fourth}, we quantify the dependence of trace
bandwidth on workload, synchronization policy and back-end choice, and derive
practical sizing rules from controlled measurements. \textbf{Fifth}, we release
the encoder together with its register sources, decoder extensions, testbenches,
formal properties and CI infrastructure, and connect it to both offline and
on-chip consumers.

Two of these deserve a word on what is and is not new. Multi-hart operation is
already provided for in the ratified specifications; what we contribute is its
evaluation on real hardware under an SMP operating system, together with
source-side process scoping. And the instrumentation path costs a single
CSR-write instruction on the software-triggered route and no target instruction
at all on the comparator-triggered one; it is low-perturbation, not free.

We are explicit about what is \emph{not} claimed. Non-intrusive hardware trace,
branch-history compression, delta-address encoding, multiple retirement and
FPGA integration are all established; N-Trace and E-Trace compliance as such is
a specification-conformance property, not a research result. \ctte{} is not the
first N-Trace hardware encoder. What is new is the combination of an open
synthesizable artifact, a dual-protocol front end that makes format cost
measurable, an evaluation in a multi-hart operating-system regime that, to the best of
our knowledge as of August 2026, no published encoder evaluation has entered,
and a verification argument whose gates are demonstrably able to fail.

\section{Background}
\label{sec:background}

\subsection{The shared ingress}

Both RISC-V trace specifications separate the core from the encoder at the same
place. Per retirement opportunity the core presents an instruction type
(\texttt{itype}: sequential, taken/non-taken direct branch, inferable or
uninferable jump, call, return, trap), the address and last-instruction size of
the block, a retire count, the privilege level and optional context, plus
exception cause and trap value. The encoder is a pure consumer of this
interface; the semantics that matter in practice are the ones integrations get
wrong---most importantly that a synchronously faulting instruction does
\emph{not} retire, so its program counter never appears in the reconstructed
stream, and that trap markers ride on non-retiring beats.

\subsection{Two output formats, one ingress}

Table~\ref{tab:formats} summarizes where the two formats actually differ.
The distinction matters for a paper: describing an encoder as ``an E-Trace
encoder with a Nexus output'' is a category error, because the compression
algorithm, the message set and the framing are all different below the shared
ingress.

\begin{table}[t]
\caption{Where E-Trace and N-Trace differ (both consume the same ITI).}
\label{tab:formats}
\centering
\footnotesize
\begin{tabular}{@{}lll@{}}
\toprule
Layer & E-Trace~\cite{riscv-etrace} & N-Trace~\cite{riscv-ntrace} \\
\midrule
Hart$\rightarrow$encoder & ITI & same ITI \\
Compression   & block / branch-map & Nexus BTM and HTM \\
Payload       & \texttt{te\_inst} packets & TCODE-identified messages \\
Framing       & E-Trace encapsulation & MDO/MSEO, self-framing \\
Control       & TCI usable & TCI integral \\
Base standard & --- & IEEE-ISTO 5001-2012~\cite{nexus5001} \\
\bottomrule
\end{tabular}
\end{table}

\subsection{Nexus messages and MDO/MSEO}

A Nexus trace is a sequence of messages, each opening with a fixed-width TCODE
followed by an ordered list of fixed- and variable-length fields. Serialization
uses two orthogonal signal groups: MDO carries message data in small chunks,
MSEO marks chunk role (data, end of variable-length field, end of message).
The stream is therefore self-framing---a decoder recovers field and message
boundaries without a length prefix, and can skip a message whose TCODE it does
not know. Variable-length fields suppress leading zeros, so an unused high bit
costs nothing. Addresses in program-trace messages are transmitted XOR-compressed
against a reference address and right-shifted by one, exploiting halfword
instruction alignment.

\subsection{The control plane}

TCI~\cite{riscv-tci} organizes a trace subsystem into typed components.
At a configured component base address, each component exposes a control
register at offset \texttt{0x000} and an implementation register at
\texttt{0x004} carrying its version and component type, followed by a
component-specific layout. This lets software identify and control a known
component uniformly; discovery of component base addresses and
interconnections remains outside TCI 1.0 and requires platform
configuration. The layout also gives vendor extensions a defined place to
live.

\subsection{What a decoder actually needs}

Reconstruction is a two-stage process. Stage one recovers typed events from the
byte stream. Stage two walks the program image from the last known address,
resolving each direct conditional branch against a history bit and each
indirect branch against a transmitted (compressed) target, until the next
synchronizing message re-anchors the position. Two consequences drive the
encoder architecture. First, the decoder must disassemble every instruction in
each traversed block, because it needs instruction sizes---so the encoder may
elide anything the decoder can derive, and nothing it cannot. Second, any
compression mechanism with cross-message state---a return-address stack, a
jump-target cache, a branch predictor---requires a \emph{bit-identical} model on
the decoder side, and therefore a defined re-convergence point after any
discontinuity. Section~\ref{sec:recovery} treats this as a first-class design
obligation rather than an implementation detail.

\section{Related Work}
\label{sec:related}

Table~\ref{tab:related} places the published hardware encoders side by side.
We report each work's own headline figures without rescaling them; the
comparability caveats are discussed in Section~\ref{sec:discussion}.

\begin{table*}[t]
\caption{Published RISC-V hardware trace encoders. Figures are as reported by
each work and are \emph{not} normalized; see Section~\ref{sec:discussion}. Our
own row gives ranges rather than single values on purpose: both quantities
depend more on the configuration and the workload than on the encoder, and
Sections~\ref{sec:eval-asic} and~\ref{sec:spectrum} measure by how much.}
\label{tab:related}
\centering
\footnotesize
\begin{tabular}{@{}p{0.18\textwidth}p{0.11\textwidth}p{0.16\textwidth}p{0.17\textwidth}p{0.19\textwidth}l@{}}
\toprule
Work & Format & Platform & Reported cost & Reported bandwidth & Public RTL \\
\midrule
Kükner et al.\ 2022~\cite{kukner2022} & RISC-V trace enc., multi-retire & 28\,nm CMOS &
  3.74\,\% area, 3.41\,\% power & down to 2.11\,bit/instr & not found \\
Hessman \& Sternvik 2024~\cite{hessman2024} & E-Trace & NOEL-V, XCKU040 &
  21\,150 LUT, $\approx$8.7\,\% of device & 0.371\,\bpi{} avg, 2.093\,\bpi{} worst & not found \\
Laghi et al.\ 2025~\cite{laghi2025} & E-Trace & CVA6/Shaheen, VCU118 &
  9.2\,\% of CVA6 subsystem & 95.1\,\% vs.\ full-opcode & yes~\cite{rvtracer} \\
Janarthanam et al.\ 2025~\cite{janarthanam2025} & N-Trace / Nexus & high-perf.\ OoO core &
  not reported & up to 2 branch records per cycle & not found \\
Karslioglu \& Akturk 2025~\cite{karslioglu2025} & N-Trace & RISC-V processor &
  1.05\,\% area & 2.06\,\bpi{} avg, 0.11\,\bpi{} best & not found \\
\midrule
\textbf{This work} & \textbf{N-Trace \emph{and} E-Trace} & \textbf{6 cores; AMD Kria KV260 + 2 open PDKs} &
  \textbf{2.1--29.1\,k LUT for single-back-end profiles; 14--15\,\% of an application-class RV64 core, \S\ref{sec:eval-asic}} &
  \textbf{0.21--4.90\,\bpi{} spectrum, \S\ref{sec:spectrum}} & \textbf{yes~\cite{ctte}} \\
\bottomrule
\end{tabular}
\end{table*}

\textbf{Multiple retirement.} Kükner et al.~\cite{kukner2022} showed in 2022
that multi-retirement encoding is practical in hardware, not merely an interface
option, reporting 3.74\,\% area and 3.41\,\% power overhead relative to a
processor core in a commercial 28\,nm process. \ctte{}'s ingress accepts one
retirement per beat, which caps ingest for wide superscalar cores; we treat this
as a stated limitation (Section~\ref{sec:discussion}) rather than claiming
novelty in this dimension.

\textbf{E-Trace for NOEL-V.} Hessman and Sternvik~\cite{hessman2024} compared
Nexus and E-Trace, chose E-Trace for its expected compression, implemented it in
VHDL, stimulated it from Spike and verified the encoding by reconstructing the
instruction flow with a third-party decoder. They report 0.371\,\bpi{} average
and 2.093\,\bpi{} worst case in delta-address mode over 36 benchmark tests,
21\,150 LUTs ($\approx$8.7\,\% of an XCKU040) and no new frequency limit for the
core, and are explicit that hardware verification remained outstanding. Their
encode--decode--reconstruct loop is the minimum bar for an encoder paper, and
we adopt it as the innermost of four layers.

\textbf{E-Trace for CVA6.} Laghi et al.~\cite{laghi2025} integrated a tracing
system---trace filter, priority, packet emitter, branch map, resynchronization
counter, register file---into the host domain of the CVA6-based Shaheen
platform~\cite{shaheen}, replicating the block-specific path per retirement lane.
They report 9.2\,\% resource overhead relative to the CVA6 subsystem
(TIP 3.2\,\%, encoder 3.5\,\%, AXI encapsulator 2.5\,\%), no critical-path
impact, and 95.1\,\% average compression against tracing every full opcode, over
14 platform tests ranging from 85.2\,\% to 99.8\,\%. The implementation is now
public~\cite{rvtracer}. It is the closest open comparator to \ctte{}, and---
because the two designs consume the same ingress---the natural indirect baseline
(Section~\ref{sec:eval-protocol}). Notably, the paper states that development of
the corresponding trace decoder was still ongoing at the time of writing, which
is the gap Section~\ref{sec:chain} addresses: an encoder is only as useful as
the chain that consumes it.

\textbf{N-Trace for high-performance cores.} Janarthanam et al.~\cite{janarthanam2025}
addressed the microarchitectural side of N-Trace: multiple branch retirements per
cycle bounded to two branch records, branch history messaging, early
determination of backpressure, VMID/ASID support, and an SRAM buffer ahead of a
slower system-memory sink. Their verification uses two oracles---field-level
comparison against a C++ reference model, and program-counter reconstruction.
We adopt the two-oracle principle and extend it with formal proofs and hardware
campaigns.

\textbf{An evaluated N-Trace tracer.} Karslioglu and Akturk~\cite{karslioglu2025}
report 2.06\,\bpi{} average, 0.11\,\bpi{} best case and 1.05\,\% area overhead.
This is the most directly comparable quantitative reference point for an N-Trace
encoder, subject to the caveats in Section~\ref{sec:discussion}.

\textbf{Encoders used for security.} Zgheib et al.~\cite{zgheib2022,zgheib2023emfi}
extend a RISC-V trace encoder to verify control-flow and code integrity and to
detect fault-injection attacks, including experimentally injected
electromagnetic faults on an Ibex-class core. Their results establish the
approach; their method also shows its present cost, in that the encoder must be
modified and the modification is visible in what it emits. \ctte{} addresses
exactly that: the same class of monitor is served by a standing on-chip event
interface, with no encoder fork and no change to the external stream
(Section~\ref{sec:rv-path}). We present that interface as a vendor extension,
never as part of either trace standard, and we evaluate no monitor here.

\textbf{Earlier RISC-V trace work.} Gamino del Río et al.~\cite{gamino2020}
modify a RISC pipeline so that instrumented trace instructions execute without
timing overhead, targeting hybrid WCET analysis. This predates both RISC-V trace
specifications and takes the opposite approach: it removes the cost of
\emph{software}-generated trace events, whereas \ctte{} consumes standardized
architectural retirement information and compresses it into an external hardware
trace stream.

\textbf{Encoders that ship with the core they observe.} Where an encoder is
delivered inside the generator or the IP package of the core it observes, it
shares a failure domain with that core: a systematic error in the shared
assumption is invisible to the pair. That is an argument for an independently
verifiable encoder regardless of how common the arrangement is, and it is the
same independence argument we apply to our own decoder cross-checks in
Section~\ref{sec:verification}.

\textbf{Software references and commercial IP.} The N-Trace working group
publishes reference C encoder and decoder code~\cite{nexrv}; open E-Trace
software encoders and decoders exist~\cite{fzi-riscv-etrace}. Commercial
E-Trace encoders and decoders also exist, including a configurable
decoder presented at RISC-V Summit Europe 2025~\cite{zak2025}; because their
RTL is not openly available, they are outside the artifact comparison.

\section{Architecture}
\label{sec:arch}

\subsection{Overview}

\ctte{} is a five-clock-domain pipeline of small, self-contained stages with
explicit ready/valid interfaces. Data flows left to right: the ingress enters in
the core's retirement clock domain, a preprocessor crosses it into the internal
processing domain, and the output stage crosses again into the sink clock
domain. A control-and-status register block, generated from SystemRDL, configures
every stage from a fifth domain, and a free-running wall clock provides the
timestamp reference.

\begin{table}[t]
\caption{Pipeline stages and responsibilities.}
\label{tab:stages}
\centering
\small
\begin{tabular}{@{}p{0.24\columnwidth}p{0.66\columnwidth}@{}}
\toprule
Stage & Responsibility \\
\midrule
Preprocessor &
  Ingress delay, clock-domain crossing, control- and data-flow filtering and
  range watchpoints, periodic / trace-enable / quota synchronization generation,
  performance counters, timestamp selection, instrumentation capture, and
  composition of the extended ingress record consumed downstream. \\
Message generator &
  Accumulates instruction count and branch history; emits synchronization,
  resource-full and trace-off correlation events; applies the compression suite.
  Protocol-agnostic. \\
Nexus formatter &
  Maps a generic message to a TCODE plus an ordered field list. \\
MDO/MSEO formatter &
  Crosses into the sink domain, serializes the field list into MDO/MSEO chunks
  and packs them into output beats with leading-zero suppression. \\
E-Trace back end &
  Alternative pair of stages producing \texttt{te\_inst} packets and their
  framing from the same generic event stream. \\
Funnel (optional) &
  Packet-aware merge of up to four output streams onto one port, never switching
  mid-packet; understands both framings. \\
\bottomrule
\end{tabular}
\end{table}

\subsection{The protocol-agnostic front end}

The design decision that distinguishes \ctte{} from the published encoders is
that everything ahead of the back end---filtering, synchronization policy,
instruction-count and history accounting, data trace, timestamps,
instrumentation---operates on a protocol-neutral event record. Only the last two
stages know a wire format. Table~\ref{tab:stages} lists the stages and what
each is responsible for; the boundary is visible in the last two rows.
Consequently:

\begin{itemize}
\item the protocol is an \emph{elaboration parameter} of the encoder module, so
      a multi-encoder SoC can instantiate different protocols side by side in
      one netlist while each instance pays for only its own back end. The
      discovery registers that report the active protocol are driven by that
      parameter rather than written by software, so they cannot lie in a mixed
      system, and the framing kind is exported as a hardware signal so a
      downstream funnel needs no second software source of truth;
\item a protocol comparison can be run with the front end held constant, which
      is what makes Section~\ref{sec:eval-protocol} a controlled experiment
      rather than a cross-paper comparison;
\item a feature added to the front end (a new synchronization cadence, a new
      filter class) is available to both formats without duplication.
\end{itemize}

\subsection{Instruction trace modes and the compression suite}

\ctte{} implements both N-Trace instruction trace modes. History mode (HTM)
accumulates one history bit per direct conditional branch and emits an indirect
branch history message at each indirect discontinuity; branch mode (BTM) emits
one message per taken direct branch and per indirect branch. The mode is
selected at run time through the WARL \texttt{trTeControl.InstMode} field, which
legalizes to HTM in builds compiled without BTM. HTM is the default because BTM
is bandwidth-inefficient; BTM exists for decoders and workflows that require
per-branch messaging.

On top of the mode, eight individually switchable compression features are
available, each with a compile-time switch and a runtime enable that resets to
zero (Table~\ref{tab:compression}). Three of them---branch prediction, jump-target
cache and repeat-instruction---take the stream outside strict N-Trace 1.0 and are
labelled as vendor extensions throughout (Section~\ref{sec:conformance}).

\begin{table}[t]
\caption{Program-trace compression features. All reset to disabled.}
\label{tab:compression}
\centering
\small
\begin{tabular}{@{}p{0.30\columnwidth}p{0.60\columnwidth}@{}}
\toprule
Feature & Effect \\
\midrule
Implicit return & Returns matching a 16-slot call/return stack carry no address. \\
Repeated history & Identical consecutive history patterns collapse into a
  resource-full message with a repeat count. \\
Wide instruction count & Raises the internal count cap from 8 to 16 bit,
  reducing count-overflow drains. The on-wire field stays variable-length, so
  this is not a wire-format change. \\
Repeat branch & Identical repeated indirect-branch-history messages collapse
  into one message with a repeat count. \\
Indirect branch history with sync & Synchronizing messages carry the pending
  history instead of pre-flushing it, replacing two messages with one. \\
Repeat instruction & Single-instruction spin loops are counted rather than
  contributing one history bit per iteration. \emph{Vendor.} \\
Jump-target cache & Indirect targets seen before are sent as a 64-entry cache
  index instead of a full address. \emph{Vendor.} \\
Branch prediction & A 512-entry predictor runs bit-identically on both sides;
  only the run length to the next mispredict is sent. \emph{Vendor.} \\
\bottomrule
\end{tabular}
\end{table}

\subsection{Control plane}

The register map is defined in SystemRDL as the single source of truth and
compiled to SystemVerilog~\cite{peakrdl}. Its layout follows
TCI~\cite{riscv-tci}: the trace encoder and the output bridge implement the
standard component layouts offset-for-offset, and three additional
components---performance counters, watchpoints, data-flow range
filters---reuse the same discovery framework in the vendor space. Optional
features that a given build does not contain read back as zero, which is the
specification-conformant way to advertise an absent option. A bus adapter ships
by default, but the CPU interface is regenerable for the other buses the
register compiler targets, so the control plane is not a porting obstacle.

\subsection{Egress}

The primary output is the standard trace bus, carrying either MDO/MSEO-framed
Nexus messages or framed \texttt{te\_inst} packets according to the instance's
protocol parameter. The selected back end is an elaboration
parameter, so a single-protocol instance contains only the required logic.
Compared with a runtime-multiplexed dual-back-end build at the same code state,
this removes $10\,060$ LUTs from a single-protocol instance.
Backpressure is legal at any time;
its consequences are the subject of Section~\ref{sec:recovery}. A separate
96-bit stream output carries \emph{uncompressed} instrumentation records for
consumers inside the fabric that must not be required to implement a Nexus
decoder. This second path is deliberately outside both standards: it is an
interface for on-chip processing, and we describe it as such rather than as a
trace format.

\subsection{Build profiles}

Every capability has its own compile switch, so a profile can remove a feature
together with its model storage and its register bits. Two consequences are
visible to a user: an omitted feature reads back as zero, and the configuration
message (Section~\ref{sec:conformance}) reports the actually compiled capability
map. Consistency between the compile switches, the register definitions and the
advertised capability bits is machine-checked in continuous integration rather
than maintained by hand.

\section{Conformance and Vendor Extensions}
\label{sec:conformance}

\subsection{Conformance posture}

We state conformance as a matrix rather than as a claim.
Table~\ref{tab:conformance} gives the condensed form; the full per-message
table, including the compile switch, the runtime control and the advertised
capability bit for each of its 43 rows, is part of the artifact and is
machine-checked against the RTL and the register description in continuous
integration.

\textbf{What the columns compare.} The three specification columns are the
\emph{published standards}---IEEE-ISTO 5001-2012, RISC-V N-Trace 1.0 and RISC-V
E-Trace 2.0, together with the Trace Control Interface where it is the normative
source. They are not any vendor's product. A commercial encoder may implement
more or less of its chosen standard than the column suggests, and we make no
statement about any specific implementation. The comparison answers ``what does
each standard define, and what does this implementation provide'', which is the
question a reader integrating against a standard actually has.

One reading has to be excluded before the table is useful. A dash means
\emph{not defined in that document}; it is a statement about a specification,
checkable against its text, and never a statement about what other
implementations can do. Where a capability sits outside all three, \ctte{}
encodes it in the reserved vendor code space and labels it as a vendor
extension; such streams require a decoder that recognizes the extension, and
with the extensions disabled the stream is byte-identical to the standard
reference.

\ctte{}'s \emph{reset} configuration produces a stream strictly inside N-Trace
1.0, with the single documented exception of the configuration message, whose
full-profile reset emits it once at trace start. Every feature that leaves that
envelope is opt-in, resets to disabled, and is labelled a vendor extension in the
documentation, in the register description and in the capability map. No
conformance certification programme exists for either specification, and we make
no certified-conformance claim.

\textbf{On the four rows that are not implemented.} We deliberately avoid a
neutral ``not implemented'' marker, because it hides the difference between an
unfinished feature and a decision. Each of the four carries its reason in the
row: one remains outside the evaluated RTL and is supported by mixed,
address-regime-dependent measurements; one is not driven at this interface
because the optional \texttt{stall} sideband towards the hart is not
implemented; one is blocked because no core we have integrated offers a
contract for the status bits it would report---so any implementation would have
to invent the semantics---and one is deferred for want of a consumer.

\subsection{The two standards optimize differently}

Table~\ref{tab:conformance} exposes something a single-standard encoder cannot
see. The two compression mechanisms we had to introduce as \emph{vendor}
extensions under N-Trace---a branch predictor whose run length replaces history
bits, and a jump-target cache index replacing a full address---are
\emph{standard} features of E-Trace, where they are format~0 subformats~0 and~1.
Conversely, repeated-history counting and a wide instruction count have no
E-Trace counterpart at all, because E-Trace has no instruction-count concept to
overflow.

The two specifications are therefore not ranked versions of one idea; their
compression envelopes are partly disjoint. An encoder that implements both
inherits the union, and---because both back ends sit behind one front end
here---the extra mechanisms cost their logic once rather than twice.

\begin{table*}[t]
\caption{Capability matrix against the \emph{published specifications}, not
against any vendor's product. $\checkmark$ = defined / implemented;
\textsf{o} = optional in that specification; \textsf{v} = vendor code space;
--- = not defined there; \textbf{V} = Accemic vendor extension;
$\blacktriangle$ = deliberately not implemented, with the reason stated in the
row. $^{\ddagger}$~In the RISC-V protocols the encoder identifies itself through the control-register model rather than through a message.}
\label{tab:conformance}
\centering
\scriptsize
\begin{tabular}{@{}p{0.30\textwidth}cccp{0.32\textwidth}@{}}
\toprule
Capability & Nexus & N-Trace & E-Trace & \ctte{} \\
\midrule
\multicolumn{5}{@{}l}{\textit{Program-flow reconstruction}}\\
Sync, branch (+sync), resource full, correlation, error &
  $\checkmark$ & $\checkmark$ & $\checkmark$ & $\checkmark$ \\
Branch-history compression &
  $\checkmark$ & $\checkmark$ (HTM) & $\checkmark$ (branch map) & $\checkmark$ \\
Per-branch messaging &
  $\checkmark$ & $\checkmark$ (BTM) & aggregated (branch map) & $\checkmark$ \\
Ownership / process context &
  $\checkmark$ & $\checkmark$ & $\checkmark$ & $\checkmark$ \\
64-bit addresses (N-Trace; E-Trace evaluated at RV32) &
  $\checkmark$ & $\checkmark$ & $\checkmark$ & $\checkmark$ (elaboration width) \\
\midrule
\multicolumn{5}{@{}l}{\textit{Bandwidth optimizations --- note the two standards differ}}\\
Implicit return &
  --- & \textsf{o} & \textsf{o} & $\checkmark$ \\
Sequential jump &
  --- & \textsf{o} & \textsf{o} & $\checkmark$ \\
Indirect branch history with sync &
  $\checkmark$ & $\checkmark$ & --- & $\checkmark$ \\
Repeat branch &
  $\checkmark$ & \textsf{o} & --- & $\checkmark$ \\
Repeated-history counting &
  --- & \textsf{o} & \textbf{---} & $\checkmark$ \\
Wide instruction count &
  --- & --- & \textbf{---} & $\checkmark$ \textbf{V} \\
Repeat instruction &
  $\checkmark$ & reserved & --- & $\checkmark$ \textbf{V} \\
Branch prediction &
  \textsf{v} & --- & $\checkmark$ (fmt.\ 0.0) & $\checkmark$ \textbf{V} \\
Jump-target cache &
  \textsf{v} & --- & $\checkmark$ (fmt.\ 0.1) & $\checkmark$ \textbf{V} \\
Virtual-address MSB elision &
  --- & \textsf{o} & \textsf{o} & $\blacktriangle$ not implemented; benefit is
  address-regime dependent (0.0\,\% for physical-address streams,
  21.2--23.2\,\% for canonical Sv39 kernel streams;
  Section~\ref{sec:discussion}) \\
\midrule
\multicolumn{5}{@{}l}{\textit{Data flow, system and control}}\\
Data trace read / write &
  $\checkmark$ & --- & $\checkmark$ & $\checkmark$ (XOR-compressed addresses) \\
Data trace with sync &
  $\checkmark$ & --- & --- & $\checkmark$ \\
Data acquisition / instrumentation payload &
  $\checkmark$ (DQM) & --- & --- & $\checkmark$ (192-bit payload) \\
Data-trace drop policy &
  \textsf{v} (TCI) & \textsf{v} (TCI) & \textsf{v} (TCI) & $\checkmark$ \\
Hart stall on backpressure (\texttt{StallEna}, instruction and data) &
  \textsf{v} (TCI) & \textsf{o} & \textsf{o} & $\blacktriangle$ optional \texttt{stall} sideband not driven; announced-loss mode used \\
Data-trace field elision &
  \textsf{v} (TCI) & \textsf{v} (TCI) & \textsf{v} (TCI) & $\blacktriangle$ deferred: decoder-visible format variant, no consumer \\
Device ID$^{\ddagger}$ &
  $\checkmark$ & --- & --- & $\checkmark$ \\
Watchpoint message &
  $\checkmark$ & --- & --- & $\checkmark$ (from the on-chip comparators) \\
Debug status &
  $\checkmark$ & --- & --- & $\blacktriangle$ no core contract for the status bits; condition for adoption documented \\
Timestamps, source ID, synchronization reasons &
  $\checkmark$ & $\checkmark$ & $\checkmark$ & $\checkmark$ \\
Announced loss with re-anchoring &
  $\checkmark$ & $\checkmark$ & $\checkmark$ & $\checkmark$; the
  \emph{bound} on re-convergence is ours and measured, not the announcement
  (Section~\ref{sec:recovery-measured}) \\
Trigger registers, overflow status, explicit sync request &
  \textsf{v} (TCI) & $\checkmark$ (TCI) & $\checkmark$ (TCI) & $\checkmark$ \\
\midrule
\multicolumn{5}{@{}l}{\textit{Encoder control registers --- functionality implied by the register models}}\\
Comparators available for filtering &
  \texttt{DC1--4} & \textsf{o} & $\checkmark$ &
  $\checkmark$ (count is an elaboration parameter, reported in the capability
  register) \\
Comparator input selectable to the context value &
  --- & \textsf{o} & $\checkmark$ &
  $\checkmark$ (\texttt{iaddr}, \texttt{context}, \texttt{tval},
  \texttt{daddr}) \\
Register holding the value to match &
  \texttt{DTSA/EA} & $\checkmark$ & $\checkmark$ &
  $\checkmark$ (\texttt{PMatch}/\texttt{SMatch}, incl.\ mask mode) \\
Filter gating from comparator, privilege, trap &
  \texttt{DTC}, \texttt{BWC} & \textsf{o} & $\checkmark$ &
  $\checkmark$ \\
Process context carried to the tool &
  OTM & TCODE 2 & Fmt.\ 3.2 &
  $\checkmark$ (both encodings, up to a 44-bit context path) \\
\midrule
\multicolumn{5}{@{}l}{\textit{Not defined in these specifications --- \ctte{} capabilities in the reserved vendor space}}\\
Software-triggered automatic message composition &
  --- & --- & --- & $\checkmark$ \textbf{V} (one write to a CSR the core need not
  implement; command classes for current PC, direct payload, data address and
  value, fetch and data thresholds, control-flow sync) \\
Event-triggered automatic message composition &
  --- & --- & --- & $\checkmark$ \textbf{V} (trigger table, $2^{n}-1$ entries at
  one lookup per cycle, own command word per entry) \\
Uncompressed event output for on-chip consumers &
  --- & --- & --- & $\checkmark$ \textbf{V} (96-bit AXI-Stream, one beat per
  event, self-describing) \\
Self-describing stream (capability map + enabled set in band) &
  --- & --- & --- & $\checkmark$ \textbf{V} \\
Per-region performance counters &
  --- & --- & --- & $\checkmark$ \textbf{V} \\
Profile scaling with register omission &
  --- & --- & --- & $\checkmark$ \textbf{V} \\
Shutdown / flush completion contract &
  --- & --- & --- & $\checkmark$ \textbf{V} \\
\bottomrule
\end{tabular}
\end{table*}

\subsection{In-band capability discovery}

A recurring practical failure mode of configurable trace encoders is decoding a
capture with the wrong assumptions: a stream produced with a compression feature
enabled is not decodable by a decoder that does not run the matching model, and
nothing in a bare byte stream says which features were active. The usual
mitigation is an out-of-band configuration file that has to survive alongside
the capture.

\ctte{} instead makes the stream self-describing. A vendor-TCODE configuration
message carries a format version, the compiled-in capability map, the runtime-enabled
subset, and the structural parameters a decoder needs to instantiate the
compression models---return-stack depth, predictor size, cache index width---taken
from the same constants the implementing modules consume, so the advertised sizes
cannot drift from the built hardware. Emission is selectable: never, once at
trace start, or additionally after every periodic synchronization. A decoder
that does not know the message still frames it correctly through MSEO and skips
it.

The verification obligation this creates is stronger than ``the message
appears''. The gate requires that a decoder invoked \emph{without} any feature
flags, configuring itself purely from the stream, produces the same reconstructed
program counters as the same decoder invoked with explicit flags, and that a
deliberately mismatched negative control diverges.

\section{Multi-Hart and Process-Scoped Tracing}
\label{sec:multihart}

\subsection{The multi-hart operating-system regime}

Every published RISC-V encoder evaluation we are aware of traces a single hart
running bare-metal or single-application code. A preemptive operating system on
several harts changes three things at once: the address space becomes 64-bit and
sparse, so address compression behaves differently; the process under
observation is descheduled, migrated and resumed, so cross-message compression
state is repeatedly invalidated by events that are not failures; and the
execution of interest is a small fraction of what the trace port carries, so the
binding constraint stops being ``can the decoder follow'' and becomes ``does the
capture buffer hold enough of the right thing''.

\subsection{Which architectural key identifies a process}

The N-Trace ownership message carries a context value, but the specification
does not say what a system should put in it, and the obvious answer is not
always available. We evaluated two cores and reached two different conclusions,
which is itself the finding.

On one core the ASID is live and usable. On the other, address-space
identifiers are configured to zero width, and the satp ASID field is tied off;
enabling it would not be a feature but a correctness bug, because that core's
TLB does not tag its entries with the ASID and the operating system would
consequently skip the TLB flush at context switch. We therefore key on
\textbf{satp.PPN}, the physical address of the page-table root. This is not a
workaround. The PPN is identical for the same process on \emph{every} hart,
which is exactly the property multi-hart process tracking needs, and it is
44~bits wide, which matches the context field. Its one operational consequence
is that it has no reset value, so the filter becomes meaningful only after the
first satp write.

A practical hardware detail proved worth encoding structurally: the context
width configured into the encoder must be at least as wide as the live key, or
the filter silently matches on a truncated identifier and names the wrong
process. The board wrapper therefore computes the connected width as
\emph{zero} unless the configured width covers the live key, so an
under-configured build degrades to ``context not reported'' rather than to
``context reported wrongly''.

\subsection{Filtering at the source, not downstream}

The encoder is armed with a context comparator, and non-matching messages are
never generated. This distinction is the whole point. A downstream filter still
pays the bandwidth and the buffer for what it later discards; a source filter
does not. In simulation the separation is exact: with the filter armed on
process~A, 204 of A's 204 instructions appear and 0 of B's 168 do, and
symmetrically for B. The unfiltered run of the same workload decodes 437
program counters against 222 and 202 for the two filtered runs, so the filtered
streams partition the unfiltered one rather than sampling it.

\subsection{Hardware evidence}

Table~\ref{tab:ownership} reports the board experiment: one guest boot, two real
Linux processes at disjoint addresses, three capture windows under equal load.

\begin{table}[t]
\caption{Process-scoped capture on hardware. Same load, same buffer; window~M
is the negative control with a deliberately mutated comparator.}
\label{tab:ownership}
\centering
\footnotesize
\begin{tabular}{@{}lrrr@{}}
\toprule
Window & decoded PCs & process A & process B \\
\midrule
A --- unfiltered            & 773\,448 & 96\,384  & 95\,380 \\
B --- armed on the A context & 763\,705 & \textbf{192\,768} & \textbf{0} \\
M --- mutated comparator     & 771\,122 & 95\,380  & 91\,364 \\
\bottomrule
\end{tabular}

\vspace{2pt}
{\footnotesize\raggedright
All three decode without a single error message. The context value identifying
process~A was recovered \emph{from} the unfiltered window; Linux does not export
the key to user space.\par}
\end{table}

Two things in that table matter more than the zero. First, window~M: a mutated
comparator restores both processes, which rules out the alternative explanation
that process~B simply was not running---the same two-sided discipline we apply
to every other gate in this work. Second, the honest way to quantify the
benefit. A byte ratio would be meaningless here because both windows
\emph{saturate} the 1\,MiB capture ring; the counter reports the buffer, not the
demand. What is measurable is observation depth: the same buffer holds
192\,768 instead of 96\,384 instructions of the target process, a factor of
2.00, which is exactly what two equally active processes should yield.

\subsection{The boundary of the mechanism}

One sentence belongs in the product documentation before anyone relies on this.
In the filtered window, 30 of 61\,120 address fields still point into the other
process's text. All 30 are indirect-branch-with-sync messages, 28 of them the
source address of an \texttt{ecall} trap. This is the documented trap exception
and not a leak---the decoder reconstructs zero instructions from them---but it
means the mechanism is an \textbf{observation filter, not an isolation
boundary}. A design that needs the latter needs a different mechanism.

\subsection{Two harts, one stream}

For multi-hart capture each hart drives its own encoder, and a packet-aware
funnel merges the streams onto one port without ever switching mid-message,
tagging each packet with its source. A decoder demultiplexes by source and
reconstructs each hart independently.

The stimulus for this case has to be chosen carefully: two harts running
identical work is the worst possible test of whether a funnel separates them,
because both reconstructions would be identical and a gate that only checks
losslessness would stay green. We therefore run the harts on disjoint address
ranges with different loop structures, and the separation gate is stated over
program-counter values rather than over the source tag---no reconstructed
address from source~0 may lie in the other hart's region, and conversely.

On that stimulus both harts reconstruct completely and independently from the
merged stream: 4\,206 of 4\,206 instructions for hart~0 and 5\,311 of 5\,311 for
hart~1, with zero error messages, and all eight separation gates pass. The
result is stated over reconstructed addresses, so it holds even if the source
tag were wrong---which is the point of stating it that way. At larger scale the
same separation holds over 3\,141\,532 instructions from one merged stream, with
zero errors and zero foreign program counters in either direction.

A negative control from that experiment is worth more than the positive one.
Wiring \emph{both} shims to the \emph{same} hart yields a decoder verdict of
``decoded OK, zero errors, both sources gapless''---a gate that checks only
losslessness would have certified a multi-core result that did not exist. This
is the same failure mode as a vacuously passing assertion, and it is why the
separation gate is stated over the addresses rather than over the source tag.

\subsection{Under an operating system}

The design boots SMP Linux on both harts, with both entering supervisor mode and
reaching a user-space login. Reconstruction there meets a limit that is worth
naming precisely, because it is not the encoder's: under Linux both harts
execute user-space code whose load addresses are randomized, so no static
listing exists for it, and instruction-level reconstruction is available for the
hart whose code we do have. The stream demonstrably carries both.

That is exactly the situation in which the ownership path earns its place,
because it does not need a listing. Over a Linux window each source reports five
distinct context values, all originating in real page-table-root writes, and the
intersection between the two sources is \textbf{empty}---reproduced in a second
window. Process separation across harts is thus established from the stream
alone, without reconstructing a single instruction of the processes involved.

Two lessons came out of getting there, and both generalize. An earlier window
\emph{could not have shown this}: each hart wrote the page-table root exactly
once, before the window opened, so the experiment was structurally incapable of
producing the evidence---the encoder was never the problem, the workload was
wrongly chosen. And the reset value of that register is zero on both harts, so a
window that catches no write does not merely fail to prove separation, it
actively suggests a shared context that does not exist. A mechanism whose
``no data'' state is indistinguishable from a positive answer needs its arming
condition checked, not assumed.

\subsection{A decoder-side finding}

Encoder work is only half of a trace path, and the operating-system regime
exposed a defect on the other half that had gone unnoticed because nothing had
ever traced across a scheduler switch. The reference decoder pinned its
return-address stack to a fixed depth and treated a mispredicted return as a
fatal condition. Any preemptively multitasking kernel was therefore decodable
only up to its first context switch. After the fix, the \emph{same} recorded
board stream yields 818\,663 instead of 34\,537 reconstructed instructions
($23.7\times$), with no error messages, and the first 34\,537 program counters
are byte-identical to the previous behaviour. The cost is measured rather than
hidden: a single-byte mutation sweep detects 25 of 40 injected corruptions where
it previously detected 27, so a strict-call-stack mode is retained and is now
the default for the simulation path.

\section{Bounded Loss and Recovery}
\label{sec:recovery}

\subsection{Why this is the hard part}

An encoder is a rate converter between a retirement stream it cannot throttle
and an output port that can stall. When the sink applies sustained backpressure,
or when a dense control-flow burst outruns the drain, internal queues overrun.
The design question is not whether loss occurs---for a fixed output bandwidth it
must, for some workload---but whether a decoder can tell that it happened, and
how quickly it regains certainty.

\ctte{}'s contract is \emph{announced loss}. A queue overrun raises an error
message whose code field records \emph{which message classes} were lost, and the
restart emits a synchronization message with a dedicated overrun reason so the
decoder re-establishes position. Nothing is dropped silently.

\subsection{The re-convergence obligation}

The subtler requirement is that announced loss is not sufficient if compression
state survives it. Every mechanism with cross-message state---the jump-target
cache mirror, the branch predictor, the implicit-return stack, the
instruction-count and history accumulators---exists in two copies, one in the
encoder and one in the decoder, and they must be bit-identical. A discontinuity
that resets one and not the other produces a decoder that is confidently wrong:
it resynchronizes, reports no error, and then mis-resolves branches.

We therefore treat re-convergence as a first-class contract with a gate per
state class: the cache mirror must be cleared on both sides at the recovery
anchor, the return stack must be empty when the trace-off correlation is
emitted, a recovery synchronization must carry a zero instruction count so that
it is a pure re-anchor, and a correlation message implies a cleared mirror and
predictor. Several of these are additionally encoded as standing assertions
inside the RTL (Section~\ref{sec:verification}) after they were first discovered
as hardware failures.

\subsection{Discontinuity taxonomy}

The same obligation applies to discontinuities that are not failures: trace
pause and resume, session stop and flush, debug and low-power entry and exit,
and mode changes. Pause closes the stream with a correlation message, suppresses
every position-less event, and re-anchors on resume with a trace-enable
synchronization; the re-anchor is required to be side-effect-free on both sides,
so that a gap does not perturb the compression models. The shutdown contract is
similarly explicit: with a stopped core and tracing still enabled, exactly one
control-flow event is permanently in flight because its successor address can
never be observed, so an ``empty'' status of zero is the correct report---closing
the stream while the core still retires is the documented order.

\subsection{Recovery bound}

The property we verify is: \emph{after an announced loss episode, the decoder is
unambiguously re-anchored no later than the next complete synchronization
message, and no compression state survives the anchor}. Section~\ref{sec:eval}
reports the observed distribution of recovery latency; the state-clearing half of
the property is proven by the invariants and formal properties of
Section~\ref{sec:verification} rather than sampled.

\section{Verification}
\label{sec:verification}

Our operating rule is that \emph{no green result counts unless the check that
produced it has been shown to be able to turn red}. Four layers implement it.

\subsection{Layer 1: round trip against an execution answer key}

A scripted core model is simultaneously the stimulus and the answer key: it
drives the ingress and records what it executed. The encoder produces a byte
stream; a reference decoder~\cite{nexrv,nexrv-ctte} reconstructs
both the program-counter sequence and a record-level export covering control
flow, memory accesses with their data values, and instrumentation. Both are
compared against what the model recorded, instruction by instruction---not
statistically. Standard-message runs use the working-group decoder as an
independent oracle; vendor-feature runs use our extended fork and are
additionally constrained by the byte-identity and disabled-state-neutrality
checks described below.

Compression features carry an additional \emph{byte-identity} obligation:
with the feature disabled the emitted stream must be byte-identical to the
historical reference, and with it enabled the reconstructed program-counter
sequence must be unchanged and the encoded byte count strictly smaller on the
feature's designated regression stimulus. This pins the wire format:
an incidental byte change in an unrelated fix turns these legs red.

\subsection{Extending the reference decoder}
\label{sec:decoder}

A round trip is only as good as the decoder closing it, and the working group's
reference decoder~\cite{nexrv} predates most of what this encoder emits. Making
it a usable oracle required a body of work we consider a contribution in its own
right, published alongside the RTL~\cite{nexrv-ctte}. Four groups of change
matter beyond our own use.

\textbf{64-bit decoding, with the width taken from the stream.} The decode state
---program counter, last address, branch source, the per-target snapshot, the
predictor index, the jump-target cache read path---is now a single
address-width type throughout, so an Sv39 kernel address survives the walk. The
\emph{printed} width follows the capture rather than a command-line flag: the
encoder announces it in band through a capability bit of the configuration
message, and the decoder adapts. We consider the absence of a flag a feature. A
capture cannot be decoded at a width its producer did not declare, so the class
of silent misinterpretation that a wrong flag produces cannot occur. Sv39 also
made two scaling defects visible that a 32-bit corpus never reached: a dense
lookup table that dereferenced a failed allocation on a huge sparse span, and a
linear record scan that cost $O(n)$ per backward branch and is now a bisection.
RV32 output is byte-identical after all of it.

\textbf{Multi-source decoding.} A funnelled stream carries interleaved messages
from several encoders. The decoder maintains a separate decode context per
source---program counter, address reference, compression models---so each hart
reconstructs independently from one merged capture, which is what makes
Section~\ref{sec:multihart} measurable at all.

\textbf{The vendor and optional message set.} Branch prediction, jump-target
cache, repeat instruction, data trace with sync and the XOR address reference
are all decoder-visible: the decoder must run a bit-identical model of the
encoder's predictor and cache, and must track the data-trace reference address.
It also consumes the configuration message to configure \emph{itself}, which is
what turns in-band capability discovery from an assertion into a mechanism.
Reserved codes it does not know are reported and skipped without perturbing the
reconstruction.

\textbf{Surviving an operating system.} The context-switch defect of
Section~\ref{sec:multihart} lived here, together with a return-stack ring
overflow that the public line still carries. Both are fixed in our fork.

One consequence must be stated plainly, because it weakens the two-oracle
argument. For the vendor features, the decoder is no longer independent of us:
we wrote both sides. Those features therefore carry the additional obligations
of Section~\ref{sec:verification}---byte-identity legs against a historical
reference and a mandatory proof that the disabled state is byte-neutral---rather
than resting on the round trip alone. Where a standard message is concerned, the
decoder logic is the working group's and the independence is real.

\subsection{Layer 2: standing invariants in the RTL}

The most expensive lessons of the hardware campaigns are encoded as concurrent
assertions inside the design source, active in every simulation of every
consumer. They state the contracts of Section~\ref{sec:recovery} directly:
no ordinary control-flow event while a resume gate holds or while tracing is
paused; an empty return stack when the trace-off correlation is emitted; a
capture fires only when an event owes one; a gap re-anchor on a call or return
retire leaves the return stack untouched; a correlation implies cleared cache
mirror and predictor count; a recovery synchronization carries a zero
instruction count; and an ``empty'' status implies an empty output queue. A
regression against any of these fires at the first simulation rather than on a
board weeks later.

\subsection{The argument at a glance}

Table~\ref{tab:verif} summarizes the four layers. We report it as a table
because the individual claims are only meaningful together: line coverage
without adversarial gates measures the testbench, formal proofs without
hardware campaigns measure the model, and campaigns without a two-sided
counter-proof measure nothing at all.

\begin{table}[t]
\caption{The verification argument, with its current numbers.}
\label{tab:verif}
\centering
\footnotesize
\begin{tabular}{@{}p{0.30\columnwidth}p{0.62\columnwidth}@{}}
\toprule
Layer & Extent \\
\midrule
Round trip &
  Every simulated run decoded and compared instruction by instruction against
  the execution answer key; standard-message runs by the working-group decoder
  as an independent oracle, vendor-feature runs by our extended fork paired
  with byte-identity and disabled-state-neutrality checks. \\
Byte identity &
  Five compression features, each proven byte-identical to the historical
  reference when disabled and lossless at strictly fewer bytes on its
  designated regression stimulus when enabled. \\
Gate battery &
  16 adversarial gates (11 robustness + 5 byte-identity legs); each encodes a
  failure mode observed on hardware and keeps the testbench that reproduced
  it. \\
In-RTL invariants &
  8 standing concurrent assertions (I1--I8), active in every simulation of
  every consumer. \\
Model checking &
  4 emission-core targets: element conservation and episode debt proven
  \emph{unbounded} by $k$-induction against an independent mirror model;
  synchronization arbitration unbounded (3 properties) plus a bounded
  non-starvation bound; message generation at BMC depth 50; the MDO/MSEO output
  stage at depth 40 with a bounded liveness guarantee at depth 120. Plus 6
  properties on the quota path. \\
Red counter-proofs &
  Every regression property fails on a checkout of the pre-fix commit
  (BMC step 6 and step 4 for the two historical defect classes). Where no
  defect commit exists, 3 of 3 injected mutations turn their property red. \\
Coverage &
  94.7\,\% line and 82.0\,\% branch coverage of the product RTL
  (5\,325 / 5\,623 lines, union over 91 coverage datasets in three build
  profiles), measured, no waivers applied. \\
Hardware campaigns &
  95/95 directed and 420 randomized soak runs, of which 411 valid, 0 failures; judged by
  machine against a reference model along 8 verdicts. \\
Simulation twin &
  343\,629 retirements with 24 natural overflow recoveries, decoding
  transitions-exact. \\
\bottomrule
\end{tabular}
\end{table}

Three encoder defect classes were found by these layers and are now locked by
them: a trap anchor, a resume-anchor race and a flush clobber. Each required
multi-day randomized board campaigns to trigger dynamically and falls to a
formal counterexample within four to six clock cycles---which is the argument
for the formal layer in one sentence.

\subsection{Layer 3: formal property gates}

Three encoder defect classes found in hardware campaigns shared one shape:
local cycle collisions with a tiny state space---phase-dependent, found
dynamically only by lucky draws. We therefore added formal property gates on a
license-free open-source route (SystemVerilog conversion~\cite{sv2v} plus
Yosys/SymbiYosys~\cite{symbiyosys}) covering the emission core:

\begin{itemize}
\item the overflow marker injector, proven by $k$-induction as a full
      functional correspondence against an independent mirror model
      (element conservation, episode debt, completion and hold semantics);
\item the synchronization arbiter, with unbounded proofs that no periodic
      synchronization may occur while a resume re-anchor is outstanding, that
      one-shot arming and dropping is correct, and that anchors occur only on
      qualified retires, plus a bounded non-starvation property;
\item the message generator, with bounded proofs that a consume beat never
      coincides with a flush-marker emission, that backpressure holds the
      message bit-stable, and that a latched flush debt is never lost;
\item the MDO/MSEO output stage, with properties covering the encoding table,
      the prohibition of the reserved dual-pin code, message conservation
      (exactly one end-of-message chunk per accepted message), beat alignment,
      stall stability and clock-domain-crossing safety, plus a bounded liveness
      guarantee.
\end{itemize}

Every regression property carries a \emph{two-sided red counter-proof}: the
property-only build passes on the fixed design and fails on a checkout of the
pre-fix commit, so the gate is demonstrably able to see the defect class it
guards. Where no historical defect commit exists---the output stage was never
broken in the field---the red side is produced by deliberate mutation (wrong
end-of-message code, suppressed full gate, suppressed end-of-message), each of
which must turn its property red. Every environment assumption is enumerated and
justified, and reachability witnesses accompany each proof so that it cannot
pass vacuously.

\subsection{Layer 4: hardware-in-the-loop campaigns}

A separate SoC integration project drives the encoder on an FPGA board through
randomized cross-term campaigns: overflow regimes from sparse to saturating
including downstream backpressure profiles, feature sets, workload mixes
(calls, loads, traps, random), synchronization modes, trace-window and
stop/flush edges, interrupt storms, one-factor bisection legs and a determinism
pair. Every run is judged by machine against a reference model of the executed
program along eight verdicts: decode integrity, loss accounting, post-recovery
losslessness, counter consistency, liveness, determinism, window semantics and
flush completeness. Draws are seeded so that any red run replays bit-identically,
and deployments are hash-verified at the target, which binds every campaign
result to the intended hardware image. Before that check existed, a silent
deploy failure could invalidate an entire campaign's attribution.

\subsection{Coverage and continuous integration}

The simulation suite is held to an explicit line-coverage target with a
documented waiver process: uncovered lines are either auto-classified as trivial
(error detectors, empty default arms) or carry an accepted waiver with a
justification and a source. The suite maps onto four CI stages: host-only checks
(decoder build and corpus, no EDA tools), the simulation gate battery including
the byte-identity legs and optional coverage, the formal properties, and the
hardware campaign on a lab runner. The first three stages are reproduced with
the published reference flow and can equally be integrated into a qualified
commercial verification environment.

\section{Evaluation}
\label{sec:eval}

\subsection{Setup}

\ctte{} has been integrated with six RISC-V cores from five suppliers, spanning
both register widths and---more importantly for an encoder---three different
ingress vocabularies (Table~\ref{tab:cores}). That breadth is the real
portability evidence. Three of the six present the standardized instruction
trace interface directly; one exposes a trace-control-interface-oriented port;
two require an adapter, in one case from a proprietary vendor trace bus. The
encoder core is identical in all six cases---every difference is absorbed by a
shim outside it, which is what makes the interface claim testable rather than
aspirational.

\begin{table}[t]
\caption{Cores integrated with \ctte{}. The ingress column is what the encoder
actually sees; adapters live outside the encoder core.}
\label{tab:cores}
\centering
\footnotesize
\begin{tabular}{@{}lllp{0.21\columnwidth}@{}}
\toprule
Core & Origin & ISA & Ingress \\
\midrule
CVA6 (32-bit) & ETH Zurich / Bologna & RV32 & standard ITI \\
CVA6 (64-bit) & ETH Zurich / Bologna & RV64, Sv39 & standard ITI \\
TGC5B & MINRES, DE & RV32 & adapter \\
EMSA5$^{\dagger}$ & Fraunhofer IPMS, DE & RV32 & standard ITI \\
Rocket & open generator & RV64 & TCI-oriented \\
MicroBlaze~V & AMD & RV32 & vendor bus, adapter \\
\bottomrule
\end{tabular}
\end{table}

Three of these run Linux; two of them do so at 64 bit.

$^{\dagger}$~EMSA5 is verified in gate-level netlist simulation with real
firmware and full decode checking, rather than on hardware; the other five
integrations carry board artifacts. Its ingress claim is the best-evidenced
single entry in the table: the wrapper wires the core's native trace ports
one-to-one, without translation.

The six integrations are not equally reproducible, and the distinction matters
for a reader who wants to rebuild rather than read. The CVA6 and Rocket
integrations are reproducible from the artifact, which carries the generator
invocation and the commit for each configuration. The remaining cores are
obtained under supplier licences that do not extend to redistribution by us, so
those integrations are \emph{reported} here but not shipped; what the artifact
provides for them is the adapter or shim on our side of the interface, which is
the part we wrote. Every quantitative result in this section---resource, timing,
area and bandwidth---is therefore produced on an open-source core, so that each
ratio has a denominator a reader can reproduce independently.

\textbf{FPGA.} Two 64-bit designs on an AMD Kria KV260 module (\texttt{xck26}), each
with one RV64 core, an encoder, and a capture sink: an application-class
in-order core (RV64IMAC, Sv39, no FPU) and an independently developed RV64IMAC
core from a different generator, chosen as an independent cross-check using a
E-Trace-oriented ingress. Both boot Linux to a user-space login. A third design
places two harts of the second core with one encoder each behind a funnel.

\textbf{ASIC.} Cell area is obtained on a license-free flow (SystemVerilog
elaboration, technology mapping, standard-cell mapping) against two open PDKs, a
130\,nm and a 130\,nm-class node, at the typical corner and \emph{without} a
timing constraint, which makes the result an area-oriented implementation
estimate rather than a timing-closed or sign-off number. Its value is not the absolute figure but the fact that
the processor cores are pushed through the \emph{same} flow, the same library
and the same memory treatment as the encoder.

We report absolute values first and ratios second, and we report payload and
wire figures separately, because the fraction of the stream consumed by framing,
synchronization and timestamps is a configuration choice rather than a property
of the encoder.

\subsection{Metrics}

We report bits per retired instruction,
\begin{equation}
\bpi = \frac{\text{emitted trace bits}}{\text{retired instructions}},
\end{equation}
in two variants: \emph{payload} \bpi{} excluding framing, and \emph{wire} \bpi{}
including MDO/MSEO or E-Trace framing and beat alignment. We additionally report
message count per thousand instructions, the message-type mix, and peak versus
mean output rate, because two encoders with the same mean \bpi{} can have very
different buffering requirements.

\subsection{Resource utilization and timing}

Table~\ref{tab:resources} reports measured FPGA figures. The full profile is the
maximum configuration; the profile mechanism removes unused feature groups
entirely. One scoping remark applies to every resource figure in this paper:
they cover the \emph{trace encoder} (and, where explicitly stated, the funnel).
The trace sink and transport---capture RAM, an external-memory path, a parallel
trace port, or a streaming link---are integration components whose cost depends
on the platform, and they are not included in any figure here.

\begin{table}[t]
\caption{Measured out-of-context FPGA resource and timing figures
(AMD \texttt{xck26}). Every row names its profile, code state and tool version;
all rows target a 5.0\,ns period. Placed designs are reported separately in
Table~\ref{tab:designs}, where the implementation constraint applies.}
\label{tab:resources}
\centering
\footnotesize
\begin{tabular}{@{}llrrrr@{}}
\toprule
Profile (OOC unless noted) & Tool & LUT & FF & BRAM & $f_{max}$ \\
\midrule
Full, N-Trace only, 08-11 & 2026.1 & 27\,331 & 20\,747 & 9.5 & 111 \\
Full, N-Trace only, 08-01 & 2026.1 & 26\,123 & 20\,022 & 9.5 & 111 \\
Full, N-Trace only, 07-20 & 2025.1 & 24\,354 & 19\,917 & 9.5 & 111 \\
Full, E-Trace only, HEAD & 2026.1 & 29\,148 & 17\,495 & 9.5 & 97.0 \\
\emph{Full, both back ends}, 08-01 & 2026.1 & \emph{36\,183} & \emph{21\,669} & \emph{9.5} & \emph{96.8} \\
Slim + gold features & 2025.1 & 3\,646 & --- & --- & $\approx$206 \\
Slim baseline & 2025.1 & 2\,517 & 2\,576 & 0 & $\approx$254 \\
Minimal slim & 2025.1 & 2\,106 & --- & --- & $\approx$255 \\
\bottomrule
\end{tabular}
\end{table}

Synthesis-tool version materially affects LUT mapping. All quantitative
comparisons in this section therefore use matching tool versions and frozen code
states; Table~\ref{tab:resources} records the complete provenance tuple for every
reported result. Results from different tool releases are retained in the
artifact for provenance but are not differenced in the paper.

The three N-Trace rows report the \emph{same} $f_{max}$, and that is not a copied
value: the worst negative slack at the 5.0\,ns target is $-4.002$\,ns in all
three runs, exactly, across two tool versions and a 12\,\% growth in logic.
\textbf{The critical path of the full profile has not moved while the design
grew}, and it is nameable: it lies in the message-framing slicer, which none of
the added feature logic touches.

The E-Trace rows answer the obvious next question and answer it differently.
\textbf{Once the E-Trace back end is present it sets the frequency}, whether or
not N-Trace stands beside it: 97.0\,MHz for the E-only profile and 96.8\,MHz for
the dual build, with the critical path in the clock-domain-crossing FIFO of
the E-Trace packetizer. The rows show a lower achieved frequency for the
E-Trace configurations and localize the path, but they do not share a common
frozen code state with the N-only rows; the exact back-end cost is isolated by
a same-revision pair in Section~\ref{sec:area-bandwidth}. That is the expected signature of
a design whose timing is set by one structural path---in this case the message
formatting chain---rather than by accumulated logic depth, and it is the reason
adding capability has so far cost area but not frequency.

Three observations. First, the profile mechanism spans an order of magnitude:
a control-flow-only slim profile is in the 2.1--2.5\,k-LUT class at
$\approx$255\,MHz, while the full profile---both protocols, all compression
features, data trace, instrumentation, timestamps, performance counters,
filters---costs about ten times that and clocks correspondingly lower. Reporting
one number for ``the encoder'' would be meaningless. Second, in the full profile
the register block is a substantial fraction of the cost, which is a direct
consequence of implementing the control-interface component layouts completely
rather than partially. Third, widening the datapath to 64-bit addresses is a
measured $+3\,897$ LUTs on the full profile, and the optional block-retirement
ingress adds $+1\,295$ LUTs / $+39$ FFs; both leave the disabled configuration
byte-identical on the reference corpus.

\subsection{What the slim profile contains, and where its area goes}

The slim rows of Table~\ref{tab:resources} are not a stripped demonstrator but
the complete N-Trace program-trace path: ingress alignment, the event composer
with its buffer stage, instruction-count and history accounting with message
selection, Nexus field assembly, MDO/MSEO slicing and output packing, overflow
announcement with re-synchronization, periodic synchronization, the timestamp
base, and a minimal control-register bank. What the profile removes is
everything else: data trace, the instrumentation paths, filters and
comparators, performance counters, and the E-Trace back end. The compression
suite is itself optional on top---the slim-gold row is the slim base plus the
full suite.

Table~\ref{tab:slimelements} breaks down where the area of such a build goes,
measured per instance on a minimal single-clock control-flow configuration
(4\,214 LUTs, 3\,483 FFs, no block RAM). The decomposition predates the packing
and width optimizations that subsequently took the slim baseline to its
2\,106-LUT figure, so the shares rather than the absolutes are the durable
result: about a third of a minimal encoder is message generation, a quarter is
the output formatting chain, a quarter is buffering, and the control registers
that dominate the full profile shrink to below a tenth once the optional
register groups are omitted.

\begin{table}[t]
\caption{Where the area of a minimal control-flow build goes. Per-instance
figures from out-of-context synthesis of a minimal single-clock configuration;
measured before the packing optimizations that produced the 2\,106-LUT slim
baseline, so the composition, not the absolute total, is the point.}
\label{tab:slimelements}
\centering
\footnotesize
\begin{tabular}{@{}lrr@{}}
\toprule
Element & LUT & FF \\
\midrule
Message generation (ICNT/HIST accounting, selection) & 1\,340 & 217 \\
MDO/MSEO slicer & 969 & 305 \\
eTIP buffer cascade (decoupling) & 805 & 1\,365 \\
Control registers (bank + bus bridge) & 333 & 291 \\
Event composer (slots, return stack, ICNT pre-stage) & $\approx$215 & $\approx$300 \\
Chunk packer + output FIFO & 152 & 79 \\
Overflow injector + resync & 137 & 36 \\
Periodic synchronization & 77 & 126 \\
Target-address sideband FIFO & 68 & 53 \\
Ingress alignment & 47 & 77 \\
Nexus formatter (field-array assembly) & 41 & 308 \\
Message buffer & 10 & 249 \\
Timestamp base & 7 & 70 \\
\midrule
Total & 4\,214$^{a}$ & 3\,483 \\
\bottomrule
\end{tabular}

\vspace{2pt}
{\footnotesize\raggedright
$^{a}$~Instance figures do not sum exactly to the total; the remainder is
top-level glue.\par}
\end{table}

\subsection{Complete 64-bit designs on hardware}

Table~\ref{tab:designs} reports the three placed-and-routed 64-bit designs. The
per-encoder figures in the two-hart row are read from the placed hierarchy, not
estimated.

\begin{table*}[t]
\caption{Placed 64-bit designs on \texttt{xck26}. The implementation constraint
is 14.062\,ns (71.11\,MHz); all rows close with zero failing endpoints, and the
achieved frequency follows from the reported slack. Encoders in the two-hart
design use the slim gold profile with a 22-bit context. Block-RAM occupancy of
the two-hart design is 56.3\,\%.}
\label{tab:designs}
\centering
\footnotesize
\begin{tabular}{@{}lrrrrrr@{}}
\toprule
Design & LUT & LUT \% & FF & CLB \% & WNS [ns] & achieved $f$ [MHz] \\
\midrule
1$\times$ application-class RV64 core + encoder & 56\,801 & 48.5 & --- & --- & $+0.636$ & 74.5 \\
1$\times$ RV64 core, 2nd generator + \emph{full-profile} encoder & 49\,504 & 42.3 & --- & 66.8 & $+2.057$ & 83.3 \\
\textbf{2}$\times$ RV64 core + \textbf{2} \emph{slim} encoders + funnel & \textbf{47\,019} & \textbf{40.2} & --- & \textbf{63.7} & $\mathbf{+3.466}$ & \textbf{94.4} \\
\quad of which encoder 0 & 4\,763 & & 4\,740 & & & \\
\quad of which encoder 1 & 4\,794 & & 4\,740 & & & \\
\bottomrule
\end{tabular}
\end{table*}

The achieved-frequency column carries a disclosure that matters for anyone
reproducing this, and it cost us a correction of our own reference point. The
in-design constraint is \textbf{14.062\,ns, that is 71.11\,MHz}, not the
13.333\,ns we had assumed: the processing-system IP drives the fabric clock from
a PLL whose achievable frequency is 71.110\,MHz, while the 13.333\,ns
\texttt{create\_clock} exists only in the out-of-context synthesis scripts and
was carried onto the in-design implementations, where it never held. All twelve
place-and-routed reports in the tree agree on the real value.

The platform's programmable-logic clock is supplied by the processing system
and, on a freshly loaded design, carries whatever the boot firmware left---on
our board 100\,MHz. Against the corrected reference that is not a modest margin:
\textbf{at 100\,MHz no design ran inside its closure}, the two-hart design
included. The achieved capabilities are 74.5 to 94.4\,MHz. We now set and read
back the fabric clock as part of loading a design rather than accepting the
value we find, we bind every reported measurement to the bitstream that produced
it by content rather than by file name, and a closure gate rejects any figure
whose bitstream cannot be linked to its timing report.

Every hardware figure reported below was then re-acquired under that regime, at
a programmed and read-back clock at or below the achieved frequency of the
corresponding design---or, where a figure could not be re-acquired because its
bitstream is no longer identifiable, removed rather than restated. Two
observations from that exercise are worth keeping. The bandwidth figure moved by
0.10\,\% across a 47\,\% span of clock frequency, which is the expected
signature of a measurement that is bounded by the workload rather than by the
port. And the campaign invalid rate fell from 44 to 9 runs out of 420, because
the slower clock produces less data per window and the capture ring wraps less
often---the invalid runs were an artifact of the measurement setup, not of the
encoder.

The two-hart row carries a methodological warning that affected our design
iteration. \textbf{On this device the binding resource is CLB occupancy, not
LUT count.} The two-hart design uses 40.2\,\% of the LUTs but 63.7\,\% of the
CLBs, while an earlier three-core design in the same family reached 99.8\,\%
CLB occupancy at 85.9\,\% LUT utilization. A utilization report taken after
synthesis does not even contain a CLB row---the figure only exists after
placement. For this design on this device, post-synthesis LUT count alone is
insufficient for a capacity claim; placed CLB occupancy is the binding metric.

The two rows in the middle of Table~\ref{tab:designs} are the more interesting
comparison, and they make the profile argument concretely rather than in
principle. \textbf{Two harts with two slim encoders are smaller and faster than
one hart with one full-profile encoder}---47\,019 against 49\,504 LUTs, 63.7
against 66.8\,\% CLB occupancy, and more timing slack. Doubling the number of
observed cores is not what costs area here; choosing the maximum feature set is.
An integrator's first lever is therefore the profile, not the core count.

Taking the two-hart design at face value, the two encoders together are
9\,557 LUTs. We report the ratio against its core system in
Section~\ref{sec:discussion} rather than here, because the denominator---a
deliberately minimal RV64 system without FPU and without a large cache
hierarchy---determines the answer more than the encoder does.

\subsection{ASIC cell area, with the cores through the identical flow}
\label{sec:eval-asic}

The ``encoder as a fraction of the core'' ratio is a figure several works in
Table~\ref{tab:related} report---with values from 1.05\,\% to 9.2\,\% where
such ratios are given---and it is the least comparable number in the field,
because each work divides its own encoder by its own core in its own
technology. We therefore push
three encoder profiles \emph{and} four processor cores through one flow, one
library and one memory treatment.

Memory handling dominates such comparisons, so we report two modes:
\emph{nomem}, in which memories are excluded from area on both sides and stated
separately as bit counts, and \emph{ff}, in which everything including memories
is mapped to standard cells. Only the first supports a statement about the
encoder.

\begin{table}[t]
\caption{ASIC cell area, \emph{nomem} mode (memories excluded on both sides and
stated as bits). Two open PDKs, identical flow, typical corner, no timing
constraint.}
\label{tab:asic}
\centering
\footnotesize
\begin{tabular}{@{}lrrr@{}}
\toprule
Design under test & mem & PDK-A & PDK-B \\
                  & [kbit] & \multicolumn{2}{c}{[k\textmu m$^2$]} \\
\midrule
\ctte{} slim gold      & 37.5  & 343   & 175 \\
\ctte{} reference slim & 38.1  & 337   & 172 \\
\ctte{} full gold      & 105.9 & 1\,471 & 914 \\
\midrule
RV32 embedded core     & 0     & 174   & 94 \\
RV64 core w/o L1       & 2.0   & 335   & 172 \\
RV64 tile incl.\ L1    & 276.7 & 990   & 492 \\
RV64 app-class core    & 1.8   & 2\,246 & 1\,219 \\
\bottomrule
\end{tabular}
\end{table}

\begin{table}[t]
\caption{Encoder area as a fraction of each core, from Table~\ref{tab:asic}.
Two PDKs, reported separately rather than averaged.}
\label{tab:asicratio}
\centering
\footnotesize
\begin{tabular}{@{}lrrrr@{}}
\toprule
Profile & / RV32 & / RV64 & / RV64 tile & / app-class \\
        & core   & w/o L1 & incl.\ L1   & RV64 \\
\midrule
slim gold      & 197 / 185\,\% & \textbf{102 / 102\,\%} & 35 / 36\,\% & 15 / 14\,\% \\
reference slim & 193 / 182\,\% & 101 / 100\,\% & 34 / 35\,\% & 15 / 14\,\% \\
full gold      & 844 / 967\,\% & 439 / 532\,\% & 149 / 186\,\% & 66 / 75\,\% \\
\bottomrule
\end{tabular}
\end{table}

Read for the pairing that actually occurs in a system, the result is
unremarkable in the best sense: against an application-class RV64 core---the
kind of core that runs the Linux workloads of this paper---a slim encoder is
\textbf{14--15\,\%} of the core's logic area in both libraries. That is of the
same order as, but not directly comparable with, the 9.2\,\% Laghi et
al.~\cite{laghi2025} report on FPGA for a comparable core, because the
technology, denominator and included logic differ; this profile carries
ownership, both instruction trace modes and the complete program-trace message
set.

The rest of the table is the methodological point, and it is why we measured the
cores at all. The \emph{same} encoder is 15\,\%, 35\,\% or 102\,\% of ``the
core'' depending only on which core is meant---an application-class RV64 core, an
RV64 tile including its L1 caches, or an RV64 core with the caches excluded. The
102\,\% figure reproduces to within one percentage point across two independently
developed libraries, so it is not measurement noise; it is what happens when a
denominator is a cacheless in-order core. Such a pairing is not representative
of the application-class systems targeted in this work; that is precisely the
point: a ratio is a statement about two things, and the published record
reports only one of them.

The same effect runs the other way. In \emph{ff} mode the cache-rich tile
denominator expands substantially, because hundreds of kilobits of cache map to
flip-flops, and the encoder fraction shrinks accordingly; those figures are not
tabulated here, so we do not use them for quantitative comparison. The
direction of the effect is the point: a denominator choice can flatter as
easily as it can alarm.

\subsection{Reference configuration and optimization scope}
\label{sec:not-optimized}

One qualification applies to every area and frequency figure in this section.
The evaluated RTL is a \textbf{technology-neutral reference configuration} whose
design priorities are functional completeness against the specifications,
configurability and verification---the priorities of
Sections~\ref{sec:conformance} and~\ref{sec:verification}. Technology- and
target-specific area, power and timing optimization is outside the scope of this
evaluation and is applied once the deployment profile and the target library are
fixed, because both determine which structures can be removed, shared or
retimed.

Two observations support that this is a stated priority rather than an excuse.
The compact output packer, the only place where we did optimize deliberately,
removed about a third of the control-flow emission path at byte-identical
output. And the critical path has not moved across three code states, which
means no timing-driven restructuring has been attempted at all. A smaller or
faster encoder is therefore entirely possible---including from us---and a
comparison of this design against an optimized one measures the optimization,
not the architecture.

The ratio those areas produce depends on what one divides by, and
Table~\ref{tab:asicratio} makes that dependence explicit rather than
resolving it in our favour: the same encoder is a very different fraction of a
core without caches, of a tile with caches, and of an application-class core.

\textbf{None of these figures is a fixed cost.} The profiles in
Table~\ref{tab:asic} are configurations of one design, not product variants:
every capability has its own compile switch, and switching it off removes its
logic, its model storage and its register bits together. Between the full and
the slim profile that is a factor of 4.3--5.3 in ASIC cell area, depending on
the library and comparison row, and---on FPGA, where
the range is wider because the register file dominates---a factor of about 12.
An integrator for whom area matters more than the last feature therefore does
not face a take-it-or-leave-it decision but a documented axis, with the cost of
each step measured rather than estimated (Section~\ref{sec:eval}, feature
costs). The full profile is the \emph{maximum} configuration, and it is the one
we measure precisely because it bounds the design from above; it is not what a
constrained system would instantiate.

A final observation with design consequences: dividing the \emph{ff}--\emph{nomem}
difference by the memory bit count shows the slim profiles at
$\approx$38 and $\approx$18\,\textmu m$^2$ per bit in the two libraries, while
the full profile costs about twice that per bit in \emph{both}. That is
structure, not noise---the extra memories of the full profile (jump-target
cache, predictor table, search trees) have multiple read ports, and a
multi-ported flip-flop array pays for each port with its own read multiplexer
tree. An SRAM macro therefore pays off twice over for the full profile.

\subsection{Feature cost}

Individual feature costs are measured by differential synthesis on small,
deterministic profiles rather than on the full profile, because full-profile
deltas below roughly $\pm 300$ LUTs are dominated by netlist repacking; a
negative delta there does not mean a feature is free.

Table~\ref{tab:featurecost} gives the ten individual switches plus the
data-acquisition bundle. It needs
\emph{two} baselines, and that requirement is itself a finding: data trace and
data acquisition cannot be measured against the compact-packer baseline at all,
because the packer refuses elaboration in those combinations. Deltas must
therefore never be differenced across the two baselines.

\begin{table}[t]
\caption{Per-feature cost, differential synthesis on two small baselines
(AMD \texttt{xck26}). Deltas are against the baseline in whose block they stand
and are not additive across features or across baselines.}
\label{tab:featurecost}
\centering
\footnotesize
\begin{tabular}{@{}lrrr@{}}
\toprule
Feature & $\Delta$LUT & $\Delta$FF & $\Delta$BRAM \\
\midrule
\multicolumn{4}{@{}l}{\textit{Baseline A, compact packer on: 3\,552 LUT, 3\,598 FF, 6.5 BRAM}}\\
Filters$^{a}$              & \textbf{+7\,822} & \textbf{+5\,085} & ---$^{c}$ \\
Branch prediction          & +618  & +56   & ---$^{c}$ \\
Jump-target cache          & +262  & +71   & ---$^{c}$ \\
Repeated history           & +86   & +91   & ---$^{c}$ \\
Implicit return            & +71   & +553  & ---$^{c}$ \\
Wide instruction count     & +66   & +29   & ---$^{c}$ \\
Repeat branch              & +31   & +162  & ---$^{c}$ \\
FIFO history               & +213  & +12   & ---$^{c}$ \\
Quota sync                 & n/m$^{d}$ & +52 & ---$^{c}$ \\
\midrule
\multicolumn{4}{@{}l}{\textit{Baseline B, compact packer off: 4\,570 LUT, 4\,221 FF}}\\
Data trace                 & +2\,859 & +3\,750 & +6 \\
Data acquisition$^{b}$     & \textbf{+8\,965} & \textbf{+10\,514} & +12.5 \\
\bottomrule
\end{tabular}

\vspace{2pt}
{\footnotesize\raggedright
$^{a}$~In the default configuration of 16 filters and 8 comparators. Both are
elaboration parameters; we quote no scaling factor because none was measured.
The flip-flop cost is accounted for: 4\,544 register bits, 89\,\% of the
measured $\Delta$FF.
$^{b}$~A bundle, not a single switch: data acquisition including the ACT blocks
and watchpoint messages. Its cost is explained structurally---five contributing
mechanisms---but not resolved into a bit budget, so we report the measurement
without a corresponding account.
$^{c}$~Not differenced in the baseline-A series.
$^{d}$~Not meaningful. Two independent measurements of this switch agree on
flip-flops ($+48$ against $+52$) and contradict each other in the \emph{sign} of
the LUT delta ($+138$ from a full-profile difference against $-27$ here). Both
magnitudes sit inside the $\pm 300$ LUT repacking band, so the LUT cost is below
the resolution of either run and we report only the flip-flop figure.\par}
\end{table}

The feature-cost data exposes a clear scaling structure: compression features
add tens to hundreds of LUTs, while observation features such as filtering, data
trace, and data acquisition form separately selectable capability blocks.
Profiles therefore let an integrator trade observation scope against area without
changing the core encoding path.

\subsection{Bandwidth}

Table~\ref{tab:bpi} lists the measured figures. All are wire figures on the
trace port, all decode without an error message, and each is stated with the
configuration that produced it, because configuration and workload both
materially affect \bpi{} and have to be reported together.

\begin{table}[t]
\caption{Measured bits per retired instruction. Wire figures, i.e.\ including
framing; the configuration associated with every point is reported explicitly.}
\label{tab:bpi}
\centering
\footnotesize
\begin{tabular}{@{}lrrr@{}}
\toprule
Capture & width / features & instr. & \bpi{} \\
\midrule
1\,MiB ring from kernel start & 64, compr.$^{b}$, sync/16 & 1\,353\,788 & \textbf{4.656} \\
Prior 32-bit baseline         & 32, none           & 1\,106\,809 & 6.139 \\
Sv39 window                   & 64, none, sync/16  & 34\,453     & 8.476 \\
Sv39 window, same capture     & 64, compr.$^{b}$, sync/16 & 25\,945 & \textbf{7.624} \\
\bottomrule
\end{tabular}

\vspace{2pt}
{\footnotesize\raggedright
$^{b}$~Compression features enabled: jump-target cache, branch predictor,
implicit return, repeat history, wide ICNT. The 64-bit boot captures run with
a dense synchronization policy---a quota of 16 instructions,
realized as one anchor per $\approx$19 instructions and counted from each
stream (71\,364 anchors over 1.35\,M instructions; 1\,399 over 25\,945)---so
these \bpi{} values are anchor-heavy and not comparable with the sync-1024
points of Table~\ref{tab:spectrum}. The uncompressed Sv39 row shares the same
configured policy; its anchor count was not separately tallied.\par}
\end{table}

Two readings follow. The compression suite has been exercised once at real
scale, on the \emph{same} capture window: 8.476 to 7.624\,\bpi{}, a reduction of
10.0\,\%. And the headline figures are not comparable to the 2.06\,\bpi{} of
Karslioglu and Akturk~\cite{karslioglu2025} or the $\approx$1.57\,\bpi{} implied
by Laghi et al.~\cite{laghi2025}---but the reason is now mechanically identified
rather than suspected, and it is not the format.

\subsection{The bandwidth spectrum}
\label{sec:spectrum}

The figures above raise the obvious question of what a \bpi{} number means at
all, and we can now answer it with a controlled experiment rather than a caveat.
Table~\ref{tab:spectrum} holds the core, the encoder build, the address filter,
the synchronization period and the window length fixed, and varies only the
\emph{program that runs}.

\begin{table*}[t]
\caption{The bandwidth spectrum: one core, one encoder build, one configuration,
four workloads. Synchronization every 1024 instructions, PC-range filtered, two
windows per point. The only difference between rows is the code being executed.
\texttt{sort} saturates this setup and is therefore reported separately, as a
transfer case, in Section~\ref{sec:spectrum-structure}.}
\label{tab:spectrum}
\centering
\footnotesize
\begin{tabular}{@{}lrrrrrrrrr@{}}
\toprule
& \multicolumn{4}{c}{\bpi{} by compression feature set}
& \multicolumn{5}{c}{structural properties of the workload} \\
\cmidrule(lr){2-5}\cmidrule(l){6-10}
Workload & none & no BP & full & best \%
& cond.\ br. & indirect & call+ret & predictable & new hist. \\
\midrule
\texttt{loop}       & 0.661 & 0.406 & \textbf{0.206} & $-69$ & 14.2\,\% & 0.00\,\% & 0.00\,\% & 97.5\,\% & 0.15\,\% \\
\texttt{coremark}   & 0.860 & 0.643 & 0.726          & $-25$ & 16.2\,\% & 0.60\,\% & 1.02\,\% & 92.2\,\% & 5.64\,\% \\
\texttt{sqlite}$^{a}$ & 2.084 & 1.710 & 1.784        & $-18$ & 12.8\,\% & 2.93\,\% & 3.50\,\% & 87.3\,\% & 5.26\,\% \\
\texttt{ptr}        & 4.627 & 4.129 & \textbf{4.900} & $-11$ & 12.1\,\% & 8.08\,\% & 8.08\,\% & 73.3\,\% & 99.44\,\% \\
\bottomrule
\end{tabular}

\vspace{2pt}
{\footnotesize\raggedright
The \% column is the best feature set against no compression.
$^{a}$~The only row with material window-to-window variation: over 15 windows
\texttt{sqlite} spans 1.77--2.08\,\bpi{} without compression
(Section~\ref{sec:spectrum-spread}). The value given is from the same window
length as the other rows.\par}
\end{table*}

\textbf{The result is a factor of 23.8}, from 0.206 to 4.900\,\bpi{}, on one
core, with one encoder netlist, in one configuration. That range is wider than
the entire spread of published encoder figures, and it is produced without
touching the encoder at all.

The consequence is not that anyone's measurement is wrong. It is that a
\bpi{} figure quoted without its workload characterizes the benchmark, not the
encoder---and that applies to our own numbers first. The 6.306\,\bpi{} directed-workload point reported in the scope
discussion (Section~\ref{sec:discussion}), the 2.06 of Karslioglu and Akturk~\cite{karslioglu2025}
and the $\approx$1.57 implied by Laghi et al.~\cite{laghi2025} all sit inside
this spectrum, and none of the three can be ranked against the others without
knowing where in it the respective workload falls. We would propose that the
field report \bpi{} with the structural properties of the workload attached, and
we give those properties in Table~\ref{tab:spectrum} for exactly that reason.

\subsection{What actually predicts the cost}
\label{sec:spectrum-structure}

The right-hand half of Table~\ref{tab:spectrum} is measured from the same
reconstructed instruction stream the \bpi{} comes from, and it identifies which
program properties drive the cost. The answer is not the obvious one.

\textbf{Conditional branch density does not explain the spread in this
four-workload data set.} It stays between 12 and 16\,\% across a range in which
the cost moves by a factor of seven. An engineer sizing a trace port by
counting branches is measuring the wrong quantity.

A corollary that surprises people is that \emph{doing nothing is expensive}. An
idle operating system costs around 3\,\bpi{} in our demonstrator---above CoreMark
and above the loop kernel---because nearly every instruction it retires belongs
to the idle loop, a timer interrupt or the scheduler, and is therefore control
flow. Trace bandwidth does not track how much work a system performs; it tracks
how often it changes direction.

\textbf{Three other properties move monotonically with the cost}: the fraction
of indirect transfers (0.00 to 8.08\,\%), the call and return density (from one
in 98 instructions to one in 12), and decreasing branch predictability (97.5
down to 73.3\,\%). The novelty of the branch history---the share of 32-branch
history words never seen before, which runs from 0.15\,\% to 99.4\,\%---separates
the endpoints strongly but is not monotonic between the two middle workloads
(5.64 against 5.26\,\%). All four are properties of the program. They can be obtained from a profile or a simulator run, before
any trace hardware exists, which makes the table usable as a sizing aid rather
than only as a result.

\textbf{And the decisive quantity is predictability, not count.} The
transfer-case workload \texttt{sort} and the \texttt{ptr} row have essentially
the same fraction of indirect transfers---7.55 and 8.08\,\%---and behave
completely differently under compression. In one, the
indirect call almost always goes to the same comparison function, the
jump-target cache hits 98.9\,\% of the time, and the compression suite removes
30\,\%. In the other the target is data-dependent, the cache hits 24.8\,\% of
the time, and only 11\,\% is left. The question to ask of one's own code is
therefore not how many indirect jumps it executes, but how many
\emph{unpredictable} ones.

\subsection{Compression features are not a one-way street}

The full compression suite is the best configuration for the loop kernel by a
wide margin, and the \emph{worst} configuration for the pointer-chasing kernel:
there it costs 6\,\% \emph{more} than no compression at all. Branch prediction
alone halves the cost at 97.5\,\% predictability, degrades it by 13\,\% at
92.2\,\%, and by 19\,\% at 73.3\,\%. The mechanism is not subtle---every
mispredicted branch costs a correction message---but the crossover point is
workload-specific, and the property that determines it is measurable on the
target code in advance.

This is the practical argument for the profile and runtime-enable structure of
Section~\ref{sec:arch}: features that help one workload and hurt another must be
individually switchable, and their reset state must be off.

\subsection{Sizing an interface: \bpi{} alone is insufficient}

What a trace port must carry is \bpi{} multiplied by the retirement rate, and
the two are not correlated. In our measurements SQLite has three times the
\bpi{} of CoreMark and nevertheless a comparable byte rate, because it retires
less than half as many instructions per second. Normalized to core clock, the
workloads in Table~\ref{tab:spectrum} span \textbf{14 to 178\,kB/s per MHz} on
the faster of our two cores. The same code on the core that retires roughly a
tenth as many instructions per cycle needs 3 to 46\,kB/s per MHz---identical
\bpi{}, an order of magnitude less bandwidth. The normalization transfers to
another clock frequency but not to another microarchitecture; the usable form is
$\text{byte/s} = \bpi \times \text{instructions/s} / 8$ with the integrator's own
retirement rate.

\subsection{The dominant term is the synchronization cadence}

The reset default was the wrong end of this curve, and it has moved. The mode
field resets to \emph{off}, so an unconfigured encoder emits no periodic
synchronization at all; the period field, however, reset to the tightest value
the hardware can express. That combination is reached by the natural sequence:
software selects ``synchronize, counted in instructions'', supplies the mode
itself, and inherits the period from reset. Measured over 100\,010 retirements
with only the period varying, the old reset cost a factor of 2.44 in payload
bytes against a period of 1024, while going beyond 1024 buys at most a further
2.5\,\% and doubles the blind window with every doubling. The reset period is
now 1024, verified at a throttled sink over 100\,008 instructions with zero
overflow announcements and byte-identical output to an unthrottled run; the same
load at the minimum period produces 790 announcements. The minimum period
remains available as an explicit stress configuration. For \emph{announced}
loss the period is irrelevant---the re-anchor is the next message---so what it
protects is the unannounced case.

In a separate stress measurement the minimum period emits a full
synchronization message carrying a complete 64-bit address every sixteen
instructions, and on a 64-bit target that single setting dominates the stream.
Changing only the period to 1024 takes a directed workload of 9\,517
instructions to 610 payload bytes in 96 messages and removes the 221--359 error
messages per run that the minimum cadence produced. This is a stress-point
comparison, not the production reset configuration.

The same conclusion arrives from the opposite direction. Running the N-Trace
\emph{reference} encoder~\cite{nexrv} over the program-counter sequence of a
real 883\,577-instruction Linux window yields 1.699\,\bpi{}, and 1.610\,\bpi{}
with a bounded return stack---i.e.\ 22\,\% below the published N-Trace figure, on
an operating-system workload rather than a benchmark. Our RTL stream for that
same window is a factor of $\approx$5.6 larger. Since both encoders implement
the same format and consume the same execution, the comparison shows that
configuration and accounting choices dominate the observed gap; it does not
isolate residual implementation differences.

The reference-encoder figures characterize the \emph{format's} potential on this
workload rather than this implementation; payload and wire figures are defined
separately in the metrics subsection and are never collapsed.

\subsection{Against the working group's own reference figures}
\label{sec:spectrum-refcheck}

The N-Trace task group publishes compression results for its reference encoder
over a public test set~\cite{ntrace-refcompression}, and comparing against them
is instructive in three separate ways.

\textbf{Their data contains the same spectrum.} At one fixed configuration
(history mode, call stack, repeat), their per-benchmark results run from
0.013 to 0.487\,\bpi{}---a factor of 37, on one encoder with one setting,
excluding an outlier they themselves set aside as unrepresentative. Their
summary lines say the same in prose: history mode beats branch mode by
3.3$\times$ \emph{on average} but ``sometimes over 8$\times$'', the call stack
gains 75\,\% ``but sometimes almost 5$\times$''. The workload term we quantify in
Table~\ref{tab:spectrum} is thus not an artifact of our setup; it is visible in
the reference data, and its authors flag it too.

\textbf{Their averages sit where our cheapest workload sits.} They report
0.229\,\bpi{} averaged over the whole set and 0.196 over the embedded-benchmark
subset; our loop kernel with the full suite measures 0.206. That agreement is
expected rather than remarkable: their set consists largely of small, loop-dominated
kernels, which is precisely the regime our loop kernel represents. What their set
does not contain is an application with a large working set and data-dependent
indirect control flow, or an operating system---the regime in which we measure
1.8 to 4.9\,\bpi{}. Our results extend theirs upward rather than contradicting
them.

\textbf{Where both measure the same benchmark, configuration and accounting
explain much of the observed gap, and we can now show it rather than argue
it.} CoreMark appears in both:
0.380\,\bpi{} in the reference measurement without compression features against
our 0.860 with none enabled. Running \emph{our} recorded CoreMark sequence
through \emph{their} encoder yields 0.506\,\bpi{}: the gap closes from a factor
of 2.3 to 1.3, and what remains is attributable to a different build of the
benchmark rather than to either encoder. Carried across all five of our
workloads, their encoder reproduces the workload axis in its own methodology, at
a factor of 15.7 between the cheapest and the most expensive against 23.8 in
ours. Two conditions travel with that number. The \texttt{sort} recording comes
from the second core while the other four come from the first, which is
immaterial to a spread but would be an unnamed variable if omitted. And the
reference encoder runs \emph{without} a return-address stack unless that option
is given: enabling it halves the two call-heavy workloads and leaves the loop
kernel bit-identical, which is at once the vacuity control of the measurement
and a warning---a single tool default here is larger than most of the
differences that cross-paper comparisons argue about. The factor of 2.3 is
not an encoder difference. Our figure is a wire measurement from hardware with
periodic synchronization every 1024 instructions and MDO/MSEO framing included;
theirs is a software encoding of a simulator-produced instruction sequence, and
the published material does not state whether framing bytes and periodic
synchronization are included. Given that changing only the synchronization
period across the measured range is worth a factor of 3.9 in wire cost on the
directed workload (Section~\ref{sec:eval}), a factor of 2.3
between a framed, synchronized hardware stream and an unframed software encoding
is what one should expect.

We draw two conclusions. First, the reference figures and ours are consistent
once the configuration is accounted for, which is a useful cross-check on both.
Second---and this is the reason we report the comparison at all---the exercise
demonstrates the paper's central methodological point on the most authoritative
N-Trace numbers available: two honest measurements of the same benchmark with
the same format differ by more than a factor of two; configuration and
accounting explain much of that gap, while this comparison does not isolate
residual implementation differences.

\subsection{How reproducible is a point in the spectrum?}
\label{sec:spectrum-spread}

A spectrum is only meaningful if its points are stable, so we measured the
window-to-window spread of every repeated series rather than assuming it.
Eight series were repeated behind Table~\ref{tab:spectrum}: the four
workloads, each under the \emph{none} and \emph{full} feature configurations,
three recorded windows per series. Six of the eight---both configurations of
pointer chasing, CoreMark and the loop kernel---agree to within
0.04--3.5\,\%: without compression, 0.04\,\% on pointer chasing, 0.12\,\% on
CoreMark and 0.76\,\% on the loop kernel; with the full feature set the widest
of the six is the loop kernel at 3.5\,\%. The two SQLite series are different
in kind: in the three-window repeat they spread by 8.6--12.7\,\%. In
addition, a longer 15-window no-compression run spans 1.77--2.08\,\bpi{}.

The explanation is more useful than the correction. \textbf{The spread does not
measure noise; it measures how uniform the application is over time.} Pointer
chasing has the highest cost and the least predictable jumps in the whole table
and is nevertheless the most reproducible workload, because it does the same
thing throughout. SQLite moves through phases---insertion, index build,
query---and a window catches whichever phase it lands in. An application-level
\bpi{} figure is therefore a property of the window as well as of the program,
which is a third variable to add to the two the rest of this section
establishes.

Across 24 measurement windows, 23 decoded successfully. One pointer-chasing,
full-suite window was rejected because of an instruction-count field; board
counters reported no trace loss. We exclude that window from the measurement
set, and the corresponding bandwidth point is supported by two independently
captured, successfully decoded windows. Because the root cause of the rejected
window remains unresolved, this point is used only as a bandwidth measurement
and not as evidence for robustness.

\subsection{Two checks on the measurement itself}

Two properties of the spectrum are worth reporting because they test the
measurement rather than the encoder.

\textbf{A repeat measurement from a different setup agrees within 4.6\,\%
across the reported ranges} (midpoints differ by 3.3\,\%). The SQLite
point was measured a second time from a separately built binary, at different
text addresses, with different filter bounds, in a different campaign on a
different day: 2.079--2.089 against 1.993--2.036\,\bpi{}. Two independent
constructions producing the same answer is the evidence that the chain measures
the program rather than itself.

The \texttt{sort} workload saturates the first setup and is therefore reported
only as a transfer case on the second core and encoder build; it is excluded
from the same-build bandwidth spectrum of Table~\ref{tab:spectrum}. Its
structural properties are properties of the program and transfer: 7.55\,\%
indirect branches, 7.60\,\% calls and returns, 73.0\,\% predictable control
flow, which places it close to \texttt{ptr} on the measured structural axes and
slightly lower in predictability.

\textbf{The ordering survives an encoder change.} Repeating the spectrum on the
second core, on a \emph{different} encoder build, reproduces the ranking
\texttt{loop} $<$ \texttt{coremark} $<$ \texttt{sqlite} $<$ \texttt{ptr} $<$
\texttt{sort} without a single inversion. The absolute values differ---the same
loop kernel costs 0.661 on one build and 0.758 on the other---and, importantly,
the difference is \emph{not} a constant factor: it is $+14.7\,\%$ on one
workload and $+0.9\,\%$ on another. That is exactly why we keep the two
spectra in separate tables and do not report a cross-core comparison: the two
board designs carry different encoder builds, so a difference between them would
be an encoder comparison wearing the costume of a core comparison.

\subsection{A negative result on an optional optimization}

The N-Trace bandwidth optimizations of chapter~9 include eliding the most
significant bits of virtual addresses. We measured what it would buy on a
32-bit corpus of seven streams (six from hardware, one from simulation):
2\,114\,730 message bytes, 319\,648 messages, 146\,323 full-address fields with
a mean of 29.78 bits, and 20\,647 XOR-compressed fields with a mean of 12.78
bits. Across all 166\,970 address fields the chunk reserve is \textbf{zero
bytes}: the existing leading-zero suppression is already exactly byte-minimal.
Enabling MSB elision on this corpus would \emph{cost} 148\,159 bytes
($+7.0\,\%$) and would benefit exactly zero fields.

The sign then reverses, and re-measuring inside timing closure showed that it
reverses with the \emph{address regime} rather than with the word width. On an
Sv39 stream, where kernel addresses carry long runs of leading ones, the same
elision \emph{saves} 23.1\,\%: 43\,731 of 47\,744 full-address fields are
compressible, and those fields account for 53.8\,\% of all message bytes.
Extending the 32-bit corpus with a stream carrying 32-bit kernel virtual
addresses reverses the sign there too, by 30\,814 bytes---so the first corpus
was not measuring the word width, it was missing a kernel range, which its own
documentation had flagged as absent. The optimization is not marginal in either
direction, and the quantity that predicts it is whether addresses carry
redundant leading bits at all.

We report this because it is the kind of result that does not otherwise get
published, and because it has a consequence for anyone reading a specification's
list of optional features: an option can be actively counterproductive in one
address regime and among the most valuable in another, so its value has to be
measured in the regime one is actually in rather than inherited from a table.
It also revises our own roadmap---MSB elision was the one chapter-9 optimization
we had deprioritized on evidence, and that evidence turns out to have been
regime-specific.

The reversal is not a property of the word width but of the address space.
Across three streams carrying canonical Sv39 kernel virtual addresses the
elision saves 21.2--23.2\,\%; on a 64-bit stream without active address
translation, carrying physical boot addresses only, it saves 0.0\,\%.
Sign-extended addresses are the mechanism, and that fourth stream is not an
outlier but the control: the mechanism the first measurement proposed as the
explanation is absent there, and so is the effect. The earlier one-stream figure
was a correlation; this is a mechanism with a measured counter-case.

\subsection{Protocol comparison at constant front end}
\label{sec:eval-protocol}

The dual back end enables a controlled back-end comparison in which the recorded
retirement stream and the complete front-end configuration are held constant
while only the selected back end changes. This removes the dominant front-end
confounders of a cross-paper comparison. It is not, however, a pure wire-format
experiment in cases where a back end realizes a configured synchronization
policy differently. We therefore verify the realized synchronization cadence
from each stream and report such deviations explicitly. The arrangement is:

\begin{center}
\begin{tabular}{@{}l@{}}
recorded ingress stream \\
\quad $\rightarrow$ \ctte{} front end (identical configuration) \\
\quad\quad $\rightarrow$ N-Trace back end $\rightarrow$ \bpi$_{\mathrm{N}}$ \\
\quad\quad $\rightarrow$ E-Trace back end $\rightarrow$ \bpi$_{\mathrm{E}}$ \\
\end{tabular}
\end{center}

\begin{table}[t]
\caption{Back-end comparison with a constant front end. The recorded ingress
and front-end configuration are held constant; the selected back end is the
primary change. Upper block: configured periodic synchronization (quota 1024),
which the two back ends realize at different cadences (on the loop kernel,
1\,950 anchors on the N arm against 440 on the E arm). Lower block: matched
rerun with periodic synchronization disabled on both arms, verified from each
stream rather than from the configuration---exactly one anchor per arm. The
ranking is unchanged in every cell. Wire \bpi{} counts protocol framing but
excludes beat-fill (idle) bytes; payload \bpi{} counts message payloads only.
Residual transport and implementation effects are reported in the text.
1.5--2.0\,M instructions per point.}
\label{tab:protocol}
\centering
\footnotesize
\begin{tabular}{@{}lrrrrrr@{}}
\toprule
& \multicolumn{3}{c}{wire \bpi{}} & \multicolumn{3}{c}{payload \bpi{}} \\
\cmidrule(lr){2-4}\cmidrule(l){5-7}
Workload & N & E & E/N & N & E & E/N \\
\midrule
\multicolumn{7}{@{}l}{\emph{Configured periodic synchronization (quota 1024):}} \\
\texttt{loop}     & 0.621 & 0.150 & \textbf{0.242} & 0.466 & 0.112 & 0.240 \\
\texttt{coremark} & 0.834 & 0.586 & 0.702 & 0.626 & 0.501 & 0.800 \\
\texttt{sqlite}   & 2.015 & 1.704 & 0.846 & 1.511 & 1.461 & 0.967 \\
\texttt{ptr}      & 4.266 & 4.241 & \textbf{0.994} & 3.200 & 3.587 & \textbf{1.121} \\
\midrule
\multicolumn{7}{@{}l}{\emph{Periodic synchronization disabled on both arms (anchor-free):}} \\
\texttt{loop}     & 0.560 & 0.141 & \textbf{0.251} & 0.420 & 0.104 & 0.247 \\
\texttt{coremark} & 0.769 & 0.556 & 0.723 & 0.577 & 0.477 & 0.827 \\
\texttt{sqlite}   & 1.942 & 1.666 & 0.858 & 1.456 & 1.430 & 0.982 \\
\texttt{ptr}      & 4.202 & 4.202 & \textbf{1.000} & 3.151 & 3.555 & \textbf{1.128} \\
\bottomrule
\end{tabular}
\end{table}

Table~\ref{tab:protocol} gives the result, and it is not the one we expected.
\textbf{The realized E-Trace back-end advantage is not a constant but a
function of the workload in these measurements}, running from a factor of
about four on the loop kernel (4.14 with the configured synchronization, 3.98
anchor-free) to \emph{parity} on pointer chasing---where, measured as payload,
the E-Trace stream is in fact 12--13\,\% wider.

The message counts, taken from the configured-synchronization arms of the upper
block, are consistent with the following mechanism. An
\emph{unpredictable} indirect jump forces \emph{both} formats to emit a message
carrying a full target address, so their cost converges. Where control flow is
predictable, N-Trace pays for each filled history with a separate resource-full
message---23\,143 of them on the loop kernel---while E-Trace carries the
corresponding control-flow information in an existing packet field. The distance between the two formats is
therefore governed by the same program property that governs the absolute
values: unpredictable indirect control flow. Two independent measurements,
Table~\ref{tab:spectrum} and Table~\ref{tab:protocol}, arrive at the same axis.

Two qualifications belong with the table, and the first of them the table now
carries as its second block rather than as a bound. Part of the loop-kernel
advantage in the upper block is saved anchors rather than saved format, because
our E-Trace packetizer realizes the configured synchronization quota only to
about a fifth on that workload. The anchor-free block repeats the comparison
with periodic synchronization switched off on both arms and moves the
loop-kernel wire ratio from 4.14 to 3.98, a change of about 3.9\,\%.
\textbf{The synchronization-cadence mismatch accounts for that margin, not for
the headline's substance}, and the arithmetic is checkable: switching anchors
off saves N-Trace 1\,949 anchors at 7.86\,B and E-Trace only 439 at 5.42\,B, so
the arm with more and costlier anchors gains more. The ranking is unchanged in
every cell. And counted in
transport \emph{beats} rather than bytes, the ranking inverts on three of the
four workloads, because our packetizer emits one byte per beat. The relation is
exact rather than approximate---beats equal wire bytes on the E arm and one
quarter of them on the N arm, on all four workloads without remainder---which is
what identifies it as a property of this implementation rather than of the
format. Anyone counting transport cost rather than payload gets the opposite
answer.

This experiment compares realized back ends under a common front end; it does not
establish the best achievable implementation of either specification, and the two
streams are not functionally interchangeable. An external cross-check is reported
only if it is completed on the same recorded ingress and counting convention.

\subsection{Bandwidth against area: the same design, both sides}
\label{sec:area-bandwidth}

Because the two back ends sit behind one front end, the comparison can be closed
on the other axis too, and here it can be closed exactly: a same-revision pair,
one commit with both arms built and only the protocol elaboration parameter
changed (out-of-context synthesis, same device and constraint as the resource
table). Under Vivado 2026.1 the realized E-Trace back end costs
$+$1\,610~LUTs, saves 1\,866 flip-flops, and closes 14.1\,MHz lower than the
N-Trace build (N: 30\,620~LUTs / 31\,241~FFs; E: 32\,230 / 29\,375). The same
pair under Vivado 2022.1.2 yields $+$3\,986~LUTs, $-$1\,763 flip-flops and
$-$26.7\,MHz: the LUT premium varies by a factor of~2.5 between tool versions
at an identical commit, so a back-end area delta without a named tool version
is not a reportable number. Each netlist was checked to contain only its own
back-end chain. The pair is built at a later code state than the dated rows of
the resource table---it carries the back-end \emph{delta}, not those rows'
absolute values---and it is an implementation-level observation, not a lower
bound for either specification.

Taken together with Table~\ref{tab:protocol}, that is a genuine engineering
trade-off rather than a superiority claim: on workloads with predictable control
flow E-Trace buys a substantial bandwidth reduction at a moderate area premium,
and on workloads dominated by unpredictable indirect jumps it buys nothing and
still costs the area. Which side of that trade a system sits on is decided by the
software it runs---the same conclusion Section~\ref{sec:spectrum} reaches from
the bandwidth side alone. To the best of our knowledge, as of August 2026, this workload-dependent
area/bandwidth trade-off has not previously been measured under a shared front
end and synthesis flow---and it is only measurable at all because both back
ends share them here.

\subsection{Robustness and recovery}

The current hardware campaign results are summarized in
Table~\ref{tab:campaigns}. The directed matrix closed all-green; the larger
randomized soak campaign closed with zero failures and a residue of nine runs
invalidated by a single instrumentation limit---the on-chip capture window
wrapping before drain---rather than by encoder behaviour, which we report
rather than exclude.

\begin{table}[t]
\caption{Hardware-in-the-loop campaign results (KV260 reference design,
machine-judged against a reference model of the executed program). Exclusions
are determined by predeclared harness limits and are not counted as encoder
verdicts.}
\label{tab:campaigns}
\centering
\small
\begin{tabular}{@{}lrrrrr@{}}
\toprule
Campaign & Attempted & Valid & Pass & Fail & Excluded$^{a}$ \\
\midrule
Directed matrix & 95 & 95 & 95 & 0 & 0 \\
Randomized soak & 420 & 411 & 411 & 0 & 9 \\
\bottomrule
\end{tabular}

\vspace{2pt}
{\footnotesize\raggedright
$^{a}$~All nine exclusions share the same predeclared harness limit: the
1\,MiB on-chip capture window wrapped before drain, overwriting the head of
the stream. This is a capture-side limit; no encoder fault is involved, and
no other exclusion cause occurred.\par}
\end{table}

A full-SoC simulation twin of the campaign workload at the board's clock profile
(343\,629 retirements, 24 natural overflow recoveries) decodes transitions-exact
and serves as the simulation-side positive control.

On the 64-bit designs, a single recorded board stream from a Linux guest
reconstructs \textbf{818\,663 instructions with zero error messages}, including
a Sv39 window in which 32\,413 of 34\,453 reconstructed program counters lie
above $2^{32}$---that is, the 64-bit path is exercised by kernel addresses
rather than only by a widened datapath in simulation.

\subsection{The demonstrator: host and guest on one device}
\label{sec:demo}

\begin{figure*}[t]
\centering
\includegraphics[width=0.86\textwidth]{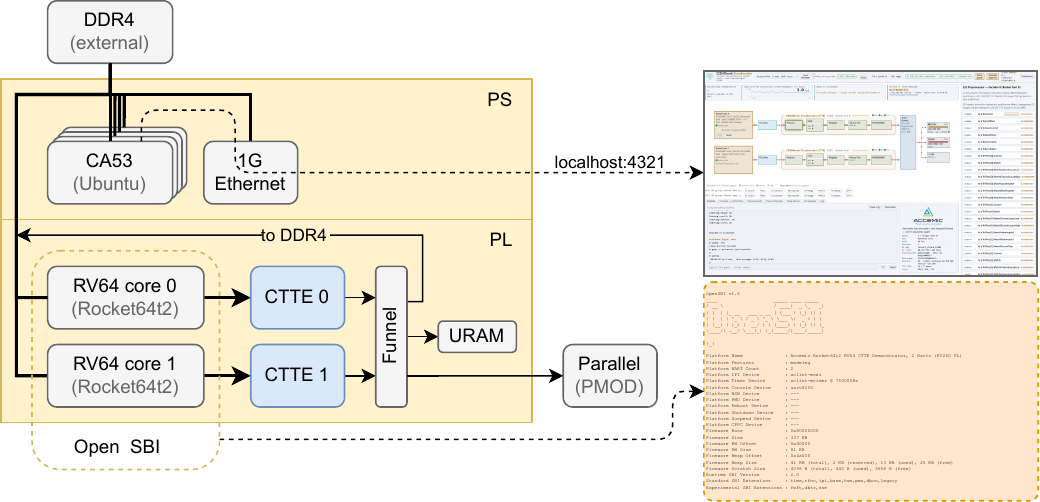}
\caption{The demonstrator by structure, and the two things it produces. In the
\emph{programmable logic} sits everything under observation: two RV64 harts,
one encoder each, a funnel merging both into one stream, and the sink---on-chip
URAM, external DDR4, or the parallel trace port on the module's PMOD connector.
The dashed region marks the harts as their firmware sees them, and the panel at
the lower right is that firmware's own boot banner---reproduced to show which
one it is, not to be read at this size. In the \emph{hard processing system}
sits everything that observes: four ARM Cortex-A53 cores under Ubuntu serve the
interface over the module's Ethernet port. That interface is the panel at the
upper right, shown here only to place it in the topology; it is reproduced
legibly, and explained, in Figure~\ref{fig:dashboard}. The two domains share a
die and neither a processor nor an operating system, which is why the interface
consumes no instructions on the traced cores.}
\label{fig:demo-setup}
\end{figure*}

The evaluation platform is worth describing, because its structure is what
makes the measurements in this section cheap enough to repeat and because it is
the artifact a reader can most directly reuse. Figure~\ref{fig:demo-setup}
gives it in one picture.

\begin{figure*}[t]
\centering
\includegraphics[width=\textwidth]{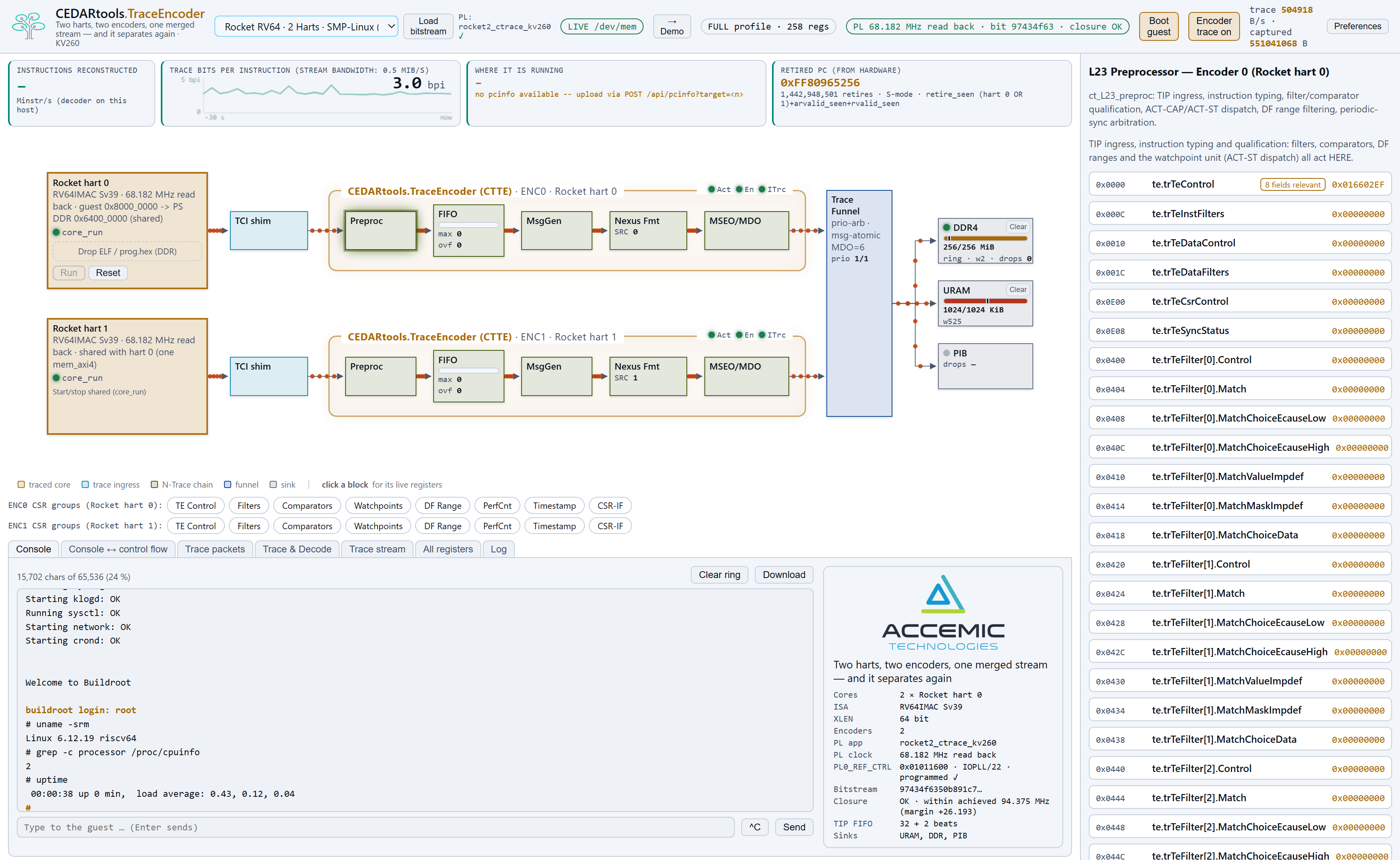}
\caption{The demonstrator. The interface is served by, and displayed on, the
device's hard processing system---four ARM cores under Ubuntu---while everything
it observes runs in the programmable logic of the same module: two RV64 harts
under SMP Linux, each with its own encoder, merged by the funnel into one stream
and separated again downstream. Observer and observed share a die but not a
processor, so the interface consumes no instructions on the traced cores;
shared interconnect, memory and trace-backpressure effects remain possible. Both encoder
pipelines are drawn stage by stage and each stage is clickable: the right pane
shows that stage's registers with their bit fields resolved to names, access
rights and live values. Bandwidth is reported as bits per retired instruction
with a rolling history, which is the quantity Section~\ref{sec:spectrum}
measures. The value shown, around 3\,\bpi{}, is the guest sitting \emph{idle}:
an idle system is an expensive one to trace, because almost every instruction it
retires belongs to the idle loop, a timer interrupt or the scheduler---that is,
to control flow. The console at the bottom is the guest's and it takes
input---the shell prompt after the Buildroot login is the traced system
answering.}
\label{fig:dashboard}
\end{figure*}

The board is a Zynq UltraScale+ MPSoC module, which carries two independent
execution domains on one device, and the split between them is what makes the
arrangement work:

\begin{itemize}
\item the \textbf{programmable logic} holds the system under observation---the
      RISC-V cores, their encoders, the funnel and the capture sink. In
      Figure~\ref{fig:dashboard} that is two RV64 harts running SMP Linux as the
      \emph{guest};
\item the \textbf{hard processing system}---four ARM Cortex-A53 cores running
      Ubuntu---holds the observer: the web server and, in the figure, the
      browser displaying it. It reaches the fabric over the on-chip
      interconnect.
\end{itemize}

Two operating systems therefore run on one die, and only one of them is being
traced. That separation is not incidental. An observer sharing the cores it
observes would be a probe effect by construction; here the user interface and
the capture control consume no instructions on the traced cores, because they
execute on different hardware---shared interconnect, memory, and
trace-backpressure effects remain possible.

The practical consequence is that there is no second machine. A conventional
arrangement needs a probe, a trace cable and a host workstation---three
components that must each be present and correctly configured before anything
can be observed at all. Here the entire observation path is on the same module
as the system under test, which is what makes the experiments in this section
cheap enough to repeat, and what makes the demonstrator fit on a conference
table.

The server exposes three things a trace experiment needs and that are usually
three separate tools. It \emph{loads} the guest---parsing an ELF image of
either width and placing it in the guest's memory window. It \emph{controls}
the encoder---the pipeline is drawn as a block diagram, and selecting a stage
lists that stage's registers with their bit fields and enumerations decoded,
live, with write access, so a configuration change is a click rather than a
recompiled test bench. And it \emph{observes}---guest console, message
statistics, a symbolized live view of the reconstructed program counter, and a
coverage map over the guest's functions.

Reconstruction for that last group uses the open reference decoder of
Section~\ref{sec:decoder}, built for the host architecture and run on the
processing system, so the whole round trip from retirement to reconstructed
program counter closes on the module without an external tool. We report this as
a convenience of the demonstrator and make no performance claim about it; the
live-processing path of Section~\ref{sec:chain} is a different mechanism, in
fabric.

Two design rules in that server are worth stating, because both were forced by
defects rather than chosen:

\textbf{The dashboard follows the board, not its own configuration file.} If
the design actually loaded into the fabric does not match the selected
scenario, the server stays in an explicitly labelled demo mode instead of
reading a register map that does not apply. Every register offset comes from the
scenario description rather than being hard-wired, because the control register
layouts of the different designs are genuinely not congruent---a server with
fixed offsets reads plausible-looking nonsense on the wrong design and reports
nothing wrong. A register a given design does not have returns ``absent'',
not zero.

\textbf{The encoder's address width is an elaboration parameter. The integration
wrapper receives the connected core's width as a mandatory parameter and refuses
elaboration if the parameter is missing or inconsistent, so an incompatible
design is rejected before a bitstream exists rather than diagnosed afterwards.
The dashboard reports the resulting encoder width. What cannot be read is
announced, not computed around.} The encoder's
address width is an elaboration parameter and therefore appears in no register.
A 64-bit core under a 32-bit encoder yields truncated addresses \emph{in
hardware}, and the interface says so in its header rather than silently
widening them again.

The message statistics from this interface also produced an early hint at the
cadence result of Section~\ref{sec:eval}: on a Linux workload, program-trace
synchronization messages accounted for 42.5\,\% of all messages, indirect
branch history for 33.2\,\% and resource-full drains for 21.3\,\%. A stream in
which more messages re-anchor the decoder than describe control flow is a
stream whose cadence, not whose format, sets its size.

\subsection{Trace as the bring-up instrument}
\label{sec:bringup}

The 64-bit Linux bring-up was not only the workload that produced the numbers
above; it was also the first setting in which the encoder had to earn its keep
as a diagnostic instrument rather than as a subject of measurement. The outcome
is the strongest practical statement we can make about it: on both cores Linux
reached a user-space login \emph{on the first attempt}, and bringing up both
systems took a single working day. We report the experience because it is the
use case that motivates the whole class of device, and because it is rarely
written down.

A kernel that stops before it initializes its console produces no output at all.
Everything the usual method offers---printing, bisecting by printing, inspecting
after the fact---requires the machine to reach a point it never reaches. The
trace stream does not have that dependency: it reports what the core executed
from the first retired instruction, whether or not any peripheral works.

During 64-bit Linux bring-up, instruction trace localized four classes of
pre-console integration failure, involving hart configuration,
atomic-interconnect signalling, ISA configuration and memory-map bounds. Three
further issues in the 64-bit peripheral path were detected in simulation before
deployment, all of the shape that boots to a console and then stops making
progress---the most expensive failure mode to diagnose without an execution
record, because the read path is correct and everything looks healthy. In each
case trace identified the failing execution point while conventional console
output was unavailable.

Both systems reached a user-space login on the first bring-up attempt, and the
complete two-core bring-up was finished within one working day. That is the
point rather than the individual defects: none of them is exotic, and each is the
kind that costs days when the only observable is a console that has not started
yet, because the search space is the whole system and every hypothesis costs a
rebuild. With an execution record the search collapses to the fix.

None of this is a quantitative result and we do not present it as one. It is,
however, the answer to the question of what a trace encoder is for, and it is
the reason the design decisions in Section~\ref{sec:chain} favour ``always
reports something truthful'' over ``reports the least possible''.

\subsection{Speculation and the absence of a correction message}

The Nexus standard defines a correction message that annuls instructions already
reported. Ratified N-Trace 1.0 does not adopt it, and we were asked whether that
is a gap. It is not, and the argument can be made positively rather than by
appeal to the specification.

Both evaluated cores execute speculatively, and on both the trace ingress is
architecturally downstream of the point where speculation is resolved: on one
core the ingress is driven from probes that report only committed instructions,
and on the other the trace port is qualified by a write-back valid signal that
already excludes replays and traps. Speculative instructions therefore never
reach the encoder, on either core, for structurally different reasons---and the
reconstructed control flow is nevertheless gapless. An encoder at this interface
can neither produce a correction message nor need one; producing one would
require a pre-retirement interface carrying cancel information. On the decoder
side we treat the reserved code as a robustness case and verify that an
unmodified decoder reports it and skips it without perturbing the
reconstruction.

\subsection{The recovery bound is a constant, not a distribution}
\label{sec:recovery-measured}

Section~\ref{sec:recovery} states the contract as ``no later than the next
complete synchronization message''. Measured over \textbf{3\,305 announced-loss
episodes} in six simulation streams, the actual distance from the error message
to the re-anchoring synchronization is \textbf{one message, four bytes and zero
instructions}---with \emph{zero} spread. There is no distribution to report.

That is a stronger property than the contract promises, and it follows from
construction rather than from a favourable sample: the encoder emits the
re-anchor as the \emph{next} message, so no other message can intervene. The six
streams span four different compression configurations---branch prediction,
implicit return, jump-target cache, and the plain overflow paths---so the
constant survives a configuration change. It is not established across board
streams or across the E-Trace back end, and we do not claim it there.

The red counter-proof is what makes the number meaningful. Evaluating the same
corpus without recognizing the synchronization message as an anchor---the
mistake a decoder makes if it treats recovery as ``wait for something
plausible''---yields a median of 28 messages, a maximum of 1\,911, and 1\,171
episodes with no anchor found at all. The measurement can distinguish a bounded
recovery from an unbounded one, which is the only reason to trust it when it
reports a bound.

\subsection{Where loss begins, and what the announcement does not tell you}
\label{sec:loss-curve}

The recovery bound says what happens after loss is announced. The remaining
question is when loss begins, and the answer is not where a mean-rate
calculation puts it.

\textbf{The quantity on the x-axis is a capacity, not an acceptance rate.} We
report \emph{sink accept capacity per unit of average encoder demand}:
accept-ready sink beats over the run, divided by the beats the encoder offered
in the lossless baseline of the same bench. The acceptance rate proper---beats
accepted per beat offered---cannot exceed 1 by construction, and is exactly
1.0000 at every clock-throttled point; it is carried as its own column in the
data set. The distinction matters because a capacity above 1 with loss below it
reads as impossible under the other reading, and the effect being reported is
precisely that.

Backpressure is applied with two levers, and both were needed. A stall injector
reproduces an Aurora-like sink that blocks in short bursts, but its elaboration
constraint caps it at about two thirds accept-ready cycles---its strongest
reachable point leaves the stream byte-identical on both benches, so it never
reaches the transition on its own. The sink clock is therefore the second
lever: a uniformly slower sink is physically the same thing as one that blocks
often, and the capacity above places both on one axis. Bench selection was
constrained the same way: a loss curve needs a baseline that is provably
lossless, and benches built to exercise overflow deliberately clock the sink far
below the core, so they begin in loss and cannot serve as a zero point.

\begin{table}[t]
\caption{Loss against sink accept capacity, 30 points on two benches
(representative rows). Capacity is accept-ready sink beats divided by the beats
offered in the same bench's lossless baseline---not an acceptance rate, see
text. Error density counts TCODE-8 messages per 1000 \emph{expected}
instructions, so the denominator is constant.}
\label{tab:loss-capacity}
\centering
\footnotesize
\begin{tabular}{@{}rrrr@{}}
\toprule
Capacity & Loss & Err./1000 instr. & Instr.\ lost per msg. \\
\midrule
\multicolumn{4}{@{}l}{\textit{Bench 1, sink-clock lever (uniform)}}\\
1.147 & 0.00\,\%  & 0     & --- \\
1.122 & 0.19\,\%  & 0.12  & 16.0 \\
1.099 & 15.63\,\% & 6.24  & 25.1 \\
1.055 & 32.51\,\% & 12.05 & 27.0 \\
0.879 & 55.19\,\% & 16.71 & 33.0 \\
0.439 & 81.91\,\% & 11.81 & 69.4 \\
\midrule
\multicolumn{4}{@{}l}{\textit{Bench 1, stall-injector lever (bursty)}}\\
1.132 & 0.00\,\%  & 0     & --- \\
1.121 & 6.94\,\%  & 2.72  & 25.5 \\
1.075 & 24.96\,\% & 8.96  & 27.9 \\
\midrule
\multicolumn{4}{@{}l}{\textit{Bench 2, sink-clock lever}}\\
1.164 & 0.00\,\%  & 0     & --- \\
1.115 & 0.46\,\%  & 0.31  & 15.2 \\
1.071 & 6.88\,\%  & 4.93  & 14.0 \\
0.892 & 29.08\,\% & 15.60 & 18.6 \\
\bottomrule
\end{tabular}
\end{table}

Table~\ref{tab:loss-capacity} gives representative points from that sweep.

\textbf{There is a threshold, and it is not at 1.} The stream stays
byte-identical to the lossless baseline down to a capacity of 1.147 on the first
bench and 1.164 on the second, and the onset sits between 1.16 and 1.11 in every
configuration measured. Below it the loss fraction does not ramp, it steps:
0.00\,\% at 1.147, 0.19\,\% at 1.122 and \textbf{15.63\,\% at 1.099}---a cliff
inside five per cent of capacity, in the same place on an independent second
bench. A sink able to absorb a tenth more than the encoder offers on average
therefore still loses trace, because the offer arrives in bursts: jump-target
cache hits in a ring and periodic synchronization, against a mean reserve that
the peaks exhaust. Sizing a trace interface to the mean rate is not sufficient,
and for this traffic type the measurement supports roughly 1.2 times the mean
rate rather than 1.0.

\textbf{At the edge, the shape of the backpressure decides; below it, only the
amount.} Well inside the loss region the two levers lie on one curve---45.8\,\%
against 45.7\,\% at a capacity of about 0.94. At the edge they separate: at
essentially the same capacity, 1.122 against 1.121, the uniform case loses
0.19\,\% and the bursty case 6.94\,\%, a factor of 36. A design sized just above
the threshold is therefore sized against the burst statistics of its
interconnect, not against its average throughput.

\textbf{No point lost trace silently, and the message count is not a measure of
how much was lost.} Across all 30 points, including the fine points immediately
below the cliff, there is no instance of missing instructions without an error
message. But the message count peaks at 276 messages for 55.19\,\% loss and
falls back to 195 for 81.91\,\%, while the instructions missing per announcement
grow from 16 to 69 (14 to 19 on the second bench). A consumer that judges trace
quality by error density therefore receives its most reassuring reading exactly
where four fifths of the control flow is absent. This is the operational form of
the property stated in Section~\ref{sec:recovery}: the stream announces
\emph{that} trace was lost, never how much, and quantifying loss requires a
reference execution---here the expected program-counter sequence of the
simulation run.

The measurement is deterministic: a point repeated in a separate working
directory reproduces character for character. Its scope is simulation, the
N-Trace back end, two ring- and branch-heavy stimuli with dense
synchronization, and stall bursts short against the buffer depth. A sink with
long blockages, credit-based backpressure or an intermediate ring buffer is not
covered, and the threshold may sit elsewhere for the E-Trace back end, whose
message and buffer sizes differ. We also report no gap \emph{count}: two
independent methods of counting gaps disagreed on the same stream---in a ring
with recurring addresses the assignment is not unique---so only directly counted
quantities are given, missing instructions and error messages.

\section{Design Rationale: The Consumer Chain}
\label{sec:chain}

A trace encoder reaches its full practical value as part of an end-to-end trace
subsystem whose consumers reconstruct and analyse its output. A trace stream has value only where something
consumes it, and the published record shows how easily that half is left open:
the closest comparable work states that its decoder is still in
development~\cite{laghi2025}, and the thesis implementation had to locate a
third-party decoder before it could verify anything at all~\cite{hessman2024}.
Several of \ctte{}'s design decisions are only explicable from the consumer
side, so we make that side explicit.

\subsection{Three consumption classes with different time budgets}

\textbf{Offline, in software.} The reference decoder of
Section~\ref{sec:decoder} reconstructs the program counter and exports
control-flow, memory-access and instrumentation records. Its time budget is
unbounded, and it is the correctness oracle for everything else. It is open
source.

\textbf{Live, in FPGA fabric.} A hardware trace processor decodes and analyses
the stream \emph{as it arrives} rather than recording it for later:
control-flow reconstruction and coverage counting are intended to run in
fabric while the target keeps running; verified throughput depends on the
implementation and link configuration. The
consequence that matters architecturally is that observation duration is not
bounded by a capture buffer, as it is for a trace streamer. Sustained operation
instead depends on the verified input rate, the selected analysis, trace
continuity and counter capacity.

\textbf{On chip, in the same fabric as the core.} The uncompressed
instrumentation stream of Section~\ref{sec:arch} serves consumers inside the
device that must not be required to implement a Nexus decoder at all. This is
the path treated in the next subsection.

One property spans all three classes and is worth stating separately. Because
the encoder is synthesizable and speaks explicit standard streaming interfaces,
the same encoder-to-decoder contract carries from simulation into FPGA
prototyping and on into SoC integration. Trace-based checks written before
silicon exists can therefore be reused during bring-up and in deployed systems
without changing the evidence format---which is what makes the bring-up
experience of Section~\ref{sec:bringup} a continuation of the verification
argument rather than a separate activity.

\subsection{What the register models leave to the integration}
\label{sec:ctxfilter}

Source-side filtering is the one capability where it is easy to overclaim,
because the mechanism is standardized and the thing that makes it useful is not.
Table~\ref{tab:conformance} shows the mechanism: all three specifications define
encoder control registers---Nexus its recommended register set, the two RISC-V
protocols the shared Trace Control Interface---and on the RISC-V side the chain
is complete, from the interface signal through a comparator that can select the
context as its input, the value written into \texttt{PMatch}, the filter, and
in-band reporting. \ctte{} implements that chain, and an encoder that does so is
conformant rather than exotic. Nexus reaches a comparable end by a different
split: its comparators are address and data comparators, and the process
identity travels in its own ownership sub-stream.

Two things are worth separating from that chain. What a given core places on the context
signal is not fixed by any of the three specifications, and where a core
supplies nothing, a usable process key has to come from somewhere; \ctte{}
derives one from the page-table root, which is an integration service rather
than a protocol capability. And the specifications say nothing about what such
filtering is worth in practice. The measurement in
Section~\ref{sec:multihart}---observation depth for a target process in a fixed
capture buffer, on a running SMP operating system---is what this paper adds to
the register definition.

\subsection{Hardware-Triggered Instrumentation without Code Modification}
\label{sec:instrumentation}

Program trace answers what the processor executed. It does not answer what a
\emph{variable} contained. Two conventional routes do, and each constrains the
experiment in a different way.

\textbf{Data trace} answers it in hardware, and \ctte{} implements it
(Section~\ref{sec:conformance}): loads and stores are reported with their
address and their value, and the program is not modified. What it costs is
bandwidth. Addresses are XOR-compressed against the previous reference, but the
value is always carried---field elision is a decoder-visible format variant and
is deferred---so a single observed store can cost more bits than a long run of
correctly predicted branches. The cost per retired instruction is then governed
by the memory-access density of the observed region rather than by its branch
structure, which is a different regime from the figures of
Section~\ref{sec:spectrum}. The practical response is to keep the observed
address windows narrow enough to hold the trace port within its budget, and that
moves the difficulty rather than removing it: the filters decide what can be
seen, and they are configured before the window is captured. One has to know
what to look for in order to see it.

\textbf{Instrumentation}---a print statement, a tracepoint, a write to a
dedicated register---makes no such advance commitment, and it can report a
derived quantity that exists in no single memory location. All of them share one
property, though: they are instructions, so they cost target cycles, perturb
timing, and change the program whose behaviour is under investigation. On a
system where the fault only appears at full speed, that is the wrong tool.

\ctte{} provides two hardware-assisted instrumentation mechanisms: one requires
a single CSR-write instruction, while the comparator-triggered path requires no
target instruction.

\textbf{Command-driven capture} costs a single CSR write. Software executes one
CSR-write instruction to the designated command CSR (\texttt{trActCapStCmd},
CSR address \texttt{0x0B10}); the encoder observes that write, including the
target CSR number, on the trace ingress interface and does the rest, assembling
a data acquisition message with a tag and up to 192 bits of payload. The target
executes one instruction, not a formatting routine, and the payload never
travels through a console, a buffer or a driver.

The part worth stating precisely is what implements that CSR: on the bus side,
nothing. The encoder does not decode it on its control interface, and the
integration's only obligation is that the write retires without an
exception---the core or its wrapper must permit the access, for example by
implementing it as a writable register with no core-side effect. What makes the
mechanism work is that the retired CSR write is visible on the instruction
trace interface, so an encoder-side feature is reached through an ordinary
core-side instruction with no bus transaction, no peripheral and no driver. The command word also carries its own routing, so the
same event can be sent to the external trace stream, to the on-chip port, to
both, or kept inside the encoder.

\textbf{Address-triggered capture requires no target instruction.} The same
message can be raised by a match rather than by an instruction: when execution
reaches an address, or when a data address is accessed, the encoder composes and
sends the record on its own. \textbf{The mechanism adds no target
instructions and requires no binary modification}---the trigger lives in the
encoder, not in the instruction stream.

Two things make this more than a pair of comparators. The trigger table is a
RAM-backed search tree with one lookup per cycle, and its depth is a
synthesis-time parameter: $2^{n}-1$ entries for a tree of $n$ levels, which is
1023 in the configuration evaluated here and reported to software in a
capability register. A design that needs more instrumentation points elaborates
a deeper tree rather than running out of them; increasing the depth increases
implementation cost.
Instrumentation points are therefore not an architectural resource to be
rationed but a table to be sized and filled. And each entry carries the same command word
as the software path, so what is captured and where it is routed is configured
per trigger rather than fixed. The table is loaded through an indirect register
pair and is locked while tracing is enabled, while read-back stays available---a
debugger can inspect the table of a running session without unlocking it.

To the best of our knowledge, as of August 2026, no other RISC-V trace encoder
offers either mechanism. Data acquisition messages are part of the Nexus
standard, so the message format is not novel; what we have not seen elsewhere is
the encoder \emph{composing} one on its own, either from a CSR write the
encoder never decodes on its bus or from a table-driven address match. Both turn
instrumentation from a code change into a configuration change, and they are
independent: a system can use the software path without a trigger table, or fill
the table without ever executing an instrumentation instruction. The
practical consequence is that a probe can be added to a running, unmodified,
production binary---including one whose source is unavailable---and removed again
by clearing a register. The advance commitment that data trace demands is not
abolished by this: one still decides before the window which addresses are of
interest. It becomes a register write between windows rather than a rebuild and
a redeployment.

\subsection{An on-chip path for runtime verification}
\label{sec:rv-path}

Using a trace encoder as the sensor for control-flow integrity checking is an
established idea: Zgheib et al.~\cite{zgheib2022,zgheib2023emfi} build a CFI and
code-integrity verifier on a RISC-V trace encoder and detect injected faults
with it, including electromagnetically induced ones. What their work also shows
is the cost of the approach as it stands: the encoder has to be
\emph{modified}, and the modification changes what the encoder emits.

\ctte{} provides that capability as a standing interface instead. A 96-bit
AXI-Stream port carries one beat per instrumentation event, self-describing:
an 8-bit identifier says what the payload means, up to three 32-bit elements
carry it, and a strobe marks which are valid. The events available on it are
the ones a runtime monitor needs---the current program counter, the program
counter pair around an exception or interrupt entry, data addresses and values
with their access type, and per-region counters---and each event is routed per
command to the external trace stream, to the on-chip port, or to both.

Three properties make this usable as a verification sensor rather than as a
debug convenience:

\begin{itemize}
\item \textbf{The payload is uncompressed.} An in-fabric consumer needs no
      Nexus decoder, no branch predictor mirror and no program image. It reads
      a beat and acts on it. That is what makes a checker implementable in the
      same fabric at the rate the events arrive.
\item \textbf{The external stream is unchanged.} Routing an event to the
      on-chip port does not perturb the standard N-Trace stream, so a monitor
      can be added to a system without invalidating an existing offline
      workflow or its evidence.
\item \textbf{Selection happens at the source.} The comparator and range
      filters that Section~\ref{sec:multihart} uses for process scoping equally
      determine which events reach the monitor, so a checker sees the region it
      is responsible for and pays no bandwidth for the rest.
\end{itemize}

We do not claim a runtime verification result here---no checker is evaluated in
this paper, and the mechanism is a vendor extension outside both trace
standards. The contribution is that the interface exists, is documented, is
covered by the same verification argument as the rest of the encoder, and is
released with it, so that a monitor is an addition to the system rather than a
fork of the encoder.

\subsection{What the chain demands of the encoder}

Each of these consumers imposes a requirement that a specification-conformance
exercise would not produce, and together they explain the encoder's less
obvious features:

\begin{itemize}
\item \textbf{A live consumer cannot repair the past.} An offline tool can
      re-read a capture and reinterpret it; a fabric pipeline processing at
      line rate cannot. Loss therefore has to be \emph{announced} in band and
      the compression models have to re-converge at a defined anchor
      (Section~\ref{sec:recovery}), because the alternative---silent loss---is
      undetectable to a consumer that only ever sees the stream once.
\item \textbf{A ring-buffer consumer needs a guaranteed anchor distance.} A
      consumer that cuts a window out of a circular buffer must know how far it
      has to read before it can decode. Synchronization cadences counted in
      emitted \emph{bytes} or \emph{messages}---rather than in instructions,
      which are only a proxy---turn that guarantee from a workload property
      into a configured number.
\item \textbf{A bandwidth-limited link makes selection worth more than
      compression.} Filtering on the process context at the source
      (Section~\ref{sec:multihart}) is what converts a system-wide bandwidth
      budget into a per-process one; no downstream stage can recover the
      capture depth that a discarded message already cost.
\item \textbf{An archived capture outlives its configuration.} In-band
      capability discovery (Section~\ref{sec:conformance}) exists because a
      capture that travels between teams, tools and years cannot rely on an
      out-of-band configuration file surviving alongside it.
\end{itemize}

\subsection{Downstream}

Beyond reconstruction, results leave the chain in standard formats---LCOV, CSV,
HTML---so that established analysis tools consume them without a proprietary
path. Concretely, timing observations feed hybrid worst-case-execution-time
analysis in AbsInt TimeWeaver; structural-coverage results obtained on hardware
are consolidated with unit-test coverage in Razorcat TESSY through its
Hyper~Coverage interface; and coverage data can be imported into CQSE Teamscale
for quality and test-gap analysis. The relevant property for this paper is not
the specific tools but the shape of the interface: an encoder whose output is
reconstructed to named coverage obligations and timing observations meets those
tools at their own input formats, and needs no privileged integration.

\emph{TimeWeaver, TESSY and Teamscale are trademarks of AbsInt Angewandte
Informatik GmbH, Razorcat Development GmbH and CQSE GmbH respectively. They are
named here solely to describe technical interoperability; there is no corporate
affiliation and no express endorsement unless separately stated.}

\section{Discussion}
\label{sec:discussion}

\subsection{Why cross-paper area ratios do not compare}

Several works in Table~\ref{tab:related} report encoder cost as a percentage
of a core, subsystem, or device; where such ratios are given, the reported
values span 1.05\,\% to 9.2\,\%. Those numbers cannot be
ranked against each other, because each divides a different encoder by a
different core in a different technology with a different treatment of memories.
Our own two slim encoders occupy 9\,557 LUTs in the two-hart system, and what
fraction that ``is'' depends entirely on the chosen system denominator; an
\emph{ff}-mode ASIC ratio from the same design looks excellent for the opposite
reason. The variable that
moves most is the denominator: an RV64 core without caches, an RV64 tile with
caches and an application-class RV64 core differ by more than a factor of seven
in Table~\ref{tab:asic}, so the same encoder is 102\,\%, 35\,\% or 15\,\% of
``the core'' depending only on where the boundary is drawn.

This is why we pushed the cores through our own flow rather than quoting theirs.
Table~\ref{tab:asicratio} is not a claim that our encoder is cheap; it is a
claim that the ratio is now defined. We would encourage the convention that a
trace encoder's cost be stated as absolute area or absolute LUTs, plus the ratio
against a \emph{named} denominator measured in the same flow.

\subsection{Sizing rules for an integrator}
\label{sec:sizing}

The spectrum of Section~\ref{sec:spectrum} is more useful as guidance than as a
result, so we state it that way. Five rules follow from the measurements, and
all five can be applied to code before any trace hardware exists.

\begin{enumerate}
\item \textbf{Budget a range, not a number.} The same encoder in the same
      configuration cost between 0.21 and 4.90\,\bpi{} on the same core. A
      \bpi{} figure without its workload describes the benchmark; this applies
      to published figures and to ours equally.
\item \textbf{Do not size by branch density.} It stayed at 12--16\,\% while the
      cost moved by a factor of seven. Count instead the indirect transfers, the
      call/return density and the branch predictability---all three moved
      monotonically in this data set. Branch-history novelty is also
      informative, but it is not monotonic at the two middle points. All are
      obtainable from a profile or a simulator run.
\item \textbf{Size by \bpi{} times retirement rate, and use your own retirement
      rate.} Our workloads spanned 14--178\,kB/s per MHz on the faster core and
      3--46 on the slower one, at identical \bpi{}. The normalization transfers
      across clock frequency, not across microarchitecture.
\item \textbf{Enable compression by measured predictability, not by default.}
      Branch prediction halved the cost at 97.5\,\% predictability and increased
      it at 92.2\,\%; on the least predictable workload the full suite was more
      expensive than no compression at all.
\item \textbf{Check the synchronization period first, and verify saturation
      through the decoder.} The period is the single largest parameter---a
      factor of 3.9 on one workload. The minimum period is the most expensive
      setting; the production reset is 1024 instructions. And when we exceeded the sustainable rate, the encoder shed trace
      while buffer-full, bus-error and drop counters all read zero; the loss was
      visible only as error messages in the \emph{decoded} stream.
\end{enumerate}

\subsection{Why cross-paper compression numbers do not compare}

The reported figures in Table~\ref{tab:related} are not on a common scale, and
converting between them is unsound without the underlying configuration.
A compression rate of 95.1\,\% against a 32-bit-per-instruction baseline
corresponds arithmetically to $\approx$1.57\,\bpi{}, but only if the same
instructions are counted, if framing, idle, alignment and synchronization are
treated identically, and if no additional trace sources are present. Published
\bpi{} figures differ in whether framing is included, in synchronization cadence,
in which optional features are enabled, in whether timestamps and context
messages are present, and in workload mix---each of which moves the result by
more than the differences being compared.

Section~\ref{sec:spectrum} turns that argument from a caveat into a measurement.
The workload term alone spans a factor of 23.8 on fixed hardware, and the
synchronization period contributes a further 3.9 on fixed workload. Both terms
are larger than the differences between any two published encoders. We therefore
report absolute values with the configuration attached, report payload and wire
separately, treat percentages as secondary, and publish the structural
properties of every workload we measure so that a reader can locate our numbers
in the spectrum rather than take them at face value.

\subsection{Scope and Integration Contracts}

We state these as contracts rather than as defects, because integrators need
them before they need a feature list.

\begin{itemize}
\item \textbf{This paper covers the trace encoder, not the trace subsystem.}
      Every resource, timing and verification figure is for the encoder and,
      where stated, the funnel. The trace sink and transport---capture RAM, an
      external-memory path, a parallel trace port or a streaming link---and the
      host-side capture infrastructure are integration components with
      platform-dependent cost, and no figure in this paper includes them.
\item \textbf{The ingress accepts one control-flow retirement per cycle.} This
      is a property of the integration as it stands today, not merely an
      untested case: the trace wrapper asserts a fatal error for any commit-port
      count other than one, the ingress shim is scalar, and the core
      configuration reduces the template's two commit ports to one. An optional
      block-retirement ingress mode exists in the encoder and is byte-neutral
      when disabled, and the evaluated core could architecturally retire two
      control-flow events per cycle---but we have no measurement of the
      combination and therefore make no claim about it in either direction.
      Kükner et al.~\cite{kukner2022} and Laghi et al.~\cite{laghi2025} remain
      the references for wide superscalar ingest.
\item \textbf{Process scoping is an observation filter, not an isolation
      boundary.} See Section~\ref{sec:multihart}: a small number of trap-source
      addresses from other processes remain visible by design. The mechanism
      reduces what is captured; it must not be relied upon to prevent
      disclosure.
\item \textbf{The process key has no reset value.} Keying on the page-table
      root means the filter becomes meaningful only after the operating system
      first writes it. Traces that must include early boot need a different
      arming strategy.
\item \textbf{Back-end comparison scope.} The controlled N-Trace / E-Trace
      comparison of Section~\ref{sec:eval-protocol} uses an RV32 ingress
      configuration; the 64-bit Linux evaluations in this work exercise the
      N-Trace path. The comparison therefore does not yet cover the
      address-width regime in which long virtual addresses dominate the stream.
      A directional figure for that regime: the same directed workload moves
      from 6.306 to 8.893\,\bpi{} on the N-Trace path when addresses widen, so
      address-field length alone is a 41\,\% effect at this synchronization
      cadence.
\item \textbf{MSB elision is not implemented.} What decides its value is the
      address regime, not the word width. Where addresses carry long runs of
      leading ones---canonical Sv39 kernel virtual addresses, and equally a
      32-bit kernel range---the elision saves: 21.2--23.2\,\% across three
      Sv39 streams, and 30\,814 bytes on a 32-bit kernel stream. Where they do
      not---physical boot addresses, whether 32-bit or 64-bit---it costs, and on
      a 64-bit stream without active address translation it saves 0.0\,\%. Our
      earlier statement that the option is counterproductive at 32 bits held only
      because the 32-bit corpus contained no kernel-range stream, a gap the
      corpus documentation had itself recorded. The figures are a model over
      recorded address fields; the encoder does not implement the elision.
\item \textbf{Byte-level determinism under overflow.} Drop positions depend on
      real clock-domain phase relations; two identical runs produce equivalent
      but not byte-identical streams. The verified property is decode
      equivalence; byte identity holds in drop-free regimes with timestamps
      disabled.
\item \textbf{Post-pause positional ambiguity.} After a pause of unknowable
      length, the first decoded segment may be positionally ambiguous against a
      full execution reference in homogeneous code. Content remains checkable;
      tooling should pair by anchor and verify transitions rather than force
      positional alignment.

\item \textbf{Core-specific ingress adapters.} Where a core exposes a
      vendor-proprietary trace bus instead of the standardized ingress, the
      adapter is pinned to the characterized core and tool version by
      construction --- such buses are typically documented as not
      backward-compatible. Adapters are therefore kept outside the encoder
      core, and their characterization gate lives with the integration
      project rather than with the encoder.
\end{itemize}

\subsection{Auditability and flow portability}

A trace encoder occupies a position of unusual trust: it reports what a
processor did, and its output is the evidence a safety or security argument
rests on. That makes independent auditability a functional property rather than
a licensing preference, particularly for regulated and long-lifecycle
programmes, where the ability to inspect, re-verify and re-implement a component
over decades is a requirement.

\ctte{}'s published artifact provides an independently auditable reference
implementation and a reproducible verification baseline. Its SystemVerilog RTL,
its assertions, its SystemRDL register sources and its build scripts are
portable across open and commercial flows, so a third party can reproduce the
evidence reported here while integrating \ctte{} into their own qualified EDA,
foundry, safety and security processes. The ASIC figures of
Section~\ref{sec:eval} were produced on open process design kits for exactly
that reason: not because a deployment must use them, but because a published
number is worth more when the flow behind it is available to the reader.

The artifact was developed in Europe against open RISC-V standards, and four of
the six cores in Table~\ref{tab:cores} originate in European research institutes
and companies. Deployment flow, target technology and qualification process
remain choices of the integrator; the published reference flow establishes
reproducibility without prescribing a production environment.

\subsection{Threats to validity}

The reference decoder used in the round-trip loop is independent of the encoder
but has been extended by us for the vendor extensions; for those features the
two-oracle property is weaker, which is why the vendor features additionally
carry byte-identity legs and a mandatory disabled-state byte-neutrality proof.
Formal properties are only as strong as their environment assumptions; these are
enumerated in the artifact and each proof carries reachability witnesses.
Campaign verdicts are only as strong as the reference model; it was validated
byte-exactly against real board captures before being used as an oracle.

The ASIC figures are unconstrained cell-area estimates rather than
timing-closed sign-off silicon-area results. Their comparative value rests
on both sides passing through the identical flow, which is the property we
control; their absolute value should not be quoted as silicon area. We also
report the library-to-library factor per design rather than a single scaling
constant, because it is not constant---across the tabulated \emph{nomem}
designs it ranges from 1.6 to 2.0, driven by cell mix.

The process-scoping result rests on identifying the context value of the target
process from the unfiltered window, because the operating system does not export
that key to user space. This is the honest procedure, but it means the
experiment demonstrates selectivity, not a turnkey workflow; a production use
needs a supported path from a process identifier to its page-table root.

\section{Conclusion}

\ctte{} is an open, synthesizable RISC-V trace encoder that consumes the
standardized instruction trace interface and drives an N-Trace/Nexus or an
E-Trace back end from one protocol-agnostic front end. It implements the
complete N-Trace 1.0 program-trace message set in both instruction trace modes,
is parametric in address width, follows the Trace Control Interface for its
register model, makes its own configuration discoverable from the stream, and
on the evaluated N-Trace simulation paths treats loss under bounded output
bandwidth as an announced, bounded and state-clearing event rather than a
silent one, with the constant one-message recovery distance established for
the six simulation streams of the recovery measurement. Its verification argument is
four-layered and, at every layer, two-sided: each regression property has a
red counter-proof, using either a pre-fix design state or an injected
mutation. The three encoder defect
classes this found had each resisted multi-day randomized board campaigns and
then fell to a formal counterexample within a handful of clock cycles---which is
the case for model checking an emission core, stated as a measurement rather
than as a preference.

Two results generalize beyond this encoder. First, tracing a preemptive
operating system on 64-bit multi-hart hardware exposes system-level conditions
that do not occur in single-hart benchmark traces---it broke our decoder before
it stressed our encoder. Filtering on the architectural process context
\emph{at the source} converts trace bandwidth from a system-wide resource into a
per-process budget, and doubled the observation depth for the target process in a
fixed capture buffer, with the exact boundary of the mechanism stated alongside.
Second, encoder-to-core area ratios are denominator-sensitive: measured in a
matched flow, a slim \ctte{} profile occupies 34--36\,\% of an RV64 tile
including its L1 caches and 14--15\,\% of an application-class RV64 core, and the
full denominator-sensitivity matrix accompanies the artifact rather than the
headline.

The resulting asset is an end-to-end trace subsystem rather than an isolated RTL
block: six demonstrated core integrations, configurable encoder profiles, a
multi-hart funnel, decoder extensions, formal and dynamic verification assets,
and interfaces to offline, live and on-chip consumers.

We publish the RTL, the register sources, the testbenches, the gate scripts and
the formal properties so that both the results and the instrument can be checked.

\subsection*{Outlook}

This encoder was not built as an end in itself. The goal behind it is
observation infrastructure for development that runs at machine speed---and
that goal explains the design decisions of this paper better than any of them
explains itself.

Generative, agentic development is already effective where its substrate is
reproducible: a container starts the same way a thousand times, and what
happened comes back as data, so a generator can propose a change, run it, read
the outcome and try again, unattended. An embedded target offers no such
ground. Cache state, interrupt latency, bus contention, supply and temperature
are physics---behaviour that no model carries in full and that shows itself
only on the real device. It cannot be assumed; it has to be observed. And when
the loop that writes the code runs at machine speed, the observation cannot
stay at human speed: assurance by review presumes an author whose intent can be
interrogated, and where software is synthesized and iterated by a generator,
that presumption weakens. The acceptance signal has to come from somewhere the
generator does not control---from the program actually executing, on the real
target, measured by an instrument outside the loop that produced the code, with
the accept-or-iterate decision taken outside that loop as well.

Read backwards, those are the requirements this paper has been answering all
along. The evidence has to be \emph{continuous and selective} rather than
sampled, because an iteration loop consumes evidence with every run and can
only afford the part of the execution it is asking about---that is what the
uncompressed event port, live in-fabric consumption and source-side process
scoping are for. Instrumentation has to move at the loop's speed: a generator
that must rebuild and redeploy to move a probe pays a compile cycle per
hypothesis, whereas the instrumentation of Section~\ref{sec:instrumentation} is
a register write between two runs, against an unmodified binary. The stream
has to stay \emph{trustworthy under overload}, because a generator scored
against evidence that quietly lost the interesting part will converge, with
confidence, on the wrong program---which is why loss is announced, its recovery
bound is measured, and the announcement's limits are stated rather than
implied. The measurement chain has to be \emph{inspectable independently of the
system that produced the code}, which is an argument for open, auditable
encoder IP whose evidence can be reproduced outside the code-generation
loop---an acceptance layer is worth little if its own correctness must be taken
on faith. And the instrument itself has to be \emph{verified to a standard at
least as high as what it certifies}, or it merely relocates the doubt---which
is why the verification argument of Section~\ref{sec:verification} is a
contribution and not an appendix.

Whether generated code becomes the norm is not ours to predict. But the
observation layer it would require begins at the trace source, it is buildable
today, and every part of it in this paper is measured, published and open to
inspection.

\appendices
\section{Artifact Availability}

The encoder RTL, SystemRDL register sources, testbenches, gate scripts, formal
property files and CI stage scripts are published at~\cite{ctte} under
CERN-OHL-S-2.0 or an Accemic commercial license, with per-file SPDX
identifiers and REUSE compliance~\cite{reuse}; scripts are ISC-licensed and
documentation is CC-BY-4.0. \ctte{} is dual-licensed: CERN-OHL-S-2.0 supports
open-hardware use, while an Accemic commercial license is available for
proprietary integration; licensing and integration enquiries are handled by
Accemic Technologies GmbH at \texttt{ctte@accemic.com}, and product
documentation is maintained at
\url{https://accemic.com/products/trace-encoder/}. The reference-decoder port used by the checking loop
is published separately~\cite{nexrv-ctte}; it derives from the N-Trace working
group's reference code~\cite{nexrv}. Simulation runs on either Verilator~\cite{verilator}
or a vendor simulator, driven by an open build tool~\cite{abcflow}; the formal
gates are reproduced with the published reference flow~\cite{sv2v,symbiyosys}
and can also be integrated into a qualified commercial verification
environment.

The processor cores used as evaluation vehicles and as area denominators are
third-party open-source designs under their own licenses. The artifact provides
the \emph{generator invocation and commit} that reproduces each configuration,
not the generated sources. For the cores obtained under supplier licences
(Table~\ref{tab:cores}), the artifact carries only our side of the
interface---the adapter or shim---since we hold no right to redistribute the
cores themselves; those integrations are reported in this paper but cannot be
rebuilt from it without a licence from the respective supplier.

One reproducibility caveat applies to the two-hart row of
Table~\ref{tab:designs}. That design was built with a revision of the trace
funnel that parses every chunk of an output beat; an earlier revision parses
only the first, which silently mis-frames a merged stream because elaboration
does not object and the symptom appears only as an unexplained decode failure
downstream. The published tree must carry the former revision for that row to be
reproducible.

The public artifact referenced in~\cite{ctte} contains
the RTL, SystemRDL sources, testbenches, formal properties, CI scripts, decoder
support, and the multi-chunk funnel revision used for the reproduced multi-hart
results.

\section*{Acknowledgment}

The encoder described in this paper was designed, implemented and verified by
Accemic Technologies GmbH. The work was carried out within the TRISTAN project,
a European research initiative to advance the RISC-V ecosystem, which provided
the collaborative context and the integration targets.

Within the TRISTAN project, Accemic Technologies GmbH has received funding from
the Chips Joint Undertaking under grant agreement No.\ 101095947, and from the
German Federal Ministry of Education and Research (BMBF) under grant
No.\ 16MEE0273. The Chips Joint Undertaking receives support from the European
Union's Horizon Europe research and innovation programme and from the
participating states.

\section*{Trademarks}

CEDARtools and Accemic are trademarks of Accemic Technologies GmbH. RISC-V,
RISC-V International and the RISC-V logos are trademarks of RISC-V
International; the marks are
used here descriptively, to name the architecture and specifications this
encoder targets. Arm, AMBA, ATB and CoreSight are trademarks or registered
trademarks of Arm Limited. AMD, Kria, Zynq, UltraScale+ and Vivado are
trademarks of Advanced Micro Devices, Inc. All other product names and brands
are the property of their respective owners and are used for identification
only.

\bibliographystyle{IEEEtran}
\bibliography{refs}

\begin{thebibliography}{10}
\providecommand{\url}[1]{#1}
\csname url@samestyle\endcsname
\providecommand{\newblock}{\relax}
\providecommand{\bibinfo}[2]{#2}
\providecommand{\BIBentrySTDinterwordspacing}{\spaceskip=0pt\relax}
\providecommand{\BIBentryALTinterwordstretchfactor}{4}
\providecommand{\BIBentryALTinterwordspacing}{\spaceskip=\fontdimen2\font plus
\BIBentryALTinterwordstretchfactor\fontdimen3\font minus
  \fontdimen4\font\relax}
\providecommand{\BIBforeignlanguage}[2]{{%
\expandafter\ifx\csname l@#1\endcsname\relax
\typeout{** WARNING: IEEEtran.bst: No hyphenation pattern has been}%
\typeout{** loaded for the language `#1'. Using the pattern for}%
\typeout{** the default language instead.}%
\else
\language=\csname l@#1\endcsname
\fi
#2}}
\providecommand{\BIBdecl}{\relax}
\BIBdecl

\bibitem{riscv-etrace}
{RISC-V International}, ``{Efficient} {Trace} for {RISC-V}, version 2.0,''
  \url{https://docs.riscv.org/reference/e-trace/index.html}, Jun. 2024,
  ratified 2022-05-05.

\bibitem{riscv-ntrace}
{RISC-V International, Nexus Trace Task Group}, ``{RISC-V} {N-Trace}
  ({Nexus-based} {Trace}) {Specification}, version 1.0,''
  \url{https://github.com/riscv-non-isa/riscv-nexus-trace}, Nov. 2024, ratified
  2024-11-21.

\bibitem{nexus5001}
{IEEE-ISTO}, ``{IEEE-ISTO} 5001-2012: The {Nexus} 5001 {Forum} {Standard} for a
  {Global} {Embedded} {Processor} {Debug} {Interface}, version 3.0.1,''
  \url{https://github.com/riscv-non-isa/riscv-nexus-trace/tree/main/docs/nexus-standard},
  2012.

\bibitem{riscv-tci}
{RISC-V International}, ``{RISC-V} {Trace} {Control} {Interface}
  {Specification}, version 1.0,''
  \url{https://docs.riscv.org/reference/trace-control-interface/index.html},
  Nov. 2024, ratified 2024-11-21.

\bibitem{kukner2022}
H.~K{\"u}kner, G.~Kaplayan, A.~Efe, and M.~A. G{\"u}lden, ``{RISC-V} processor
  trace encoder with multiple instructions retirement support,'' in \emph{2022
  IFIP/IEEE 30th International Conference on Very Large Scale Integration
  (VLSI-SoC)}.\hskip 1em plus 0.5em minus 0.4em\relax Patras, Greece: IEEE,
  2022, {IEEE} Xplore document 9939596.

\bibitem{hessman2024}
M.~Hessman and O.~Sternvik, ``Implementation of a trace encoder for {NOEL-V}
  {RISC-V} processors,'' Master's thesis, Chalmers University of Technology and
  University of Gothenburg, Gothenburg, Sweden, 2024,
  \url{https://odr.chalmers.se/handle/20.500.12380/308623}.

\bibitem{laghi2025}
U.~Laghi, S.~Manoni, E.~Parisi, and A.~Bartolini, ``Efficient trace for
  {RISC-V}: Design, evaluation, and integration in {CVA6},'' in \emph{RISC-V
  Summit Europe}, Paris, France, 2025, arXiv:2504.01972 [cs.AR].

\bibitem{rvtracer}
{PULP Platform}, ``{rv\_tracer} --- {RISC-V} {E-Trace} encoder {RTL},''
  \url{https://github.com/pulp-platform/rv_tracer}, 2025.

\bibitem{janarthanam2025}
S.~Janarthanam, R.~Behl, S.~R, and H.~P. Oleti, ``Implementing and verifying
  {RISC-V} {Nexus} {Trace} compliant trace encoder for high performance
  cores,'' in \emph{DVCon India}, 2025, {Paper} Session 3C: RISC-V/Processor.

\bibitem{karslioglu2025}
O.~Karslioglu and I.~Akturk, ``Design and evaluation of an {N-Trace} compliant
  hardware tracer for {RISC-V} processors,'' in \emph{2025 IEEE 43rd
  International Conference on Computer Design (ICCD)}.\hskip 1em plus 0.5em
  minus 0.4em\relax Richardson, TX, USA: IEEE, 2025, p. 666, {ISBN}
  979-8-3315-0347-5.

\bibitem{ctte}
{Accemic Technologies GmbH}, ``{CEDARtools.TraceEncoder} ({CTTE}) --- trace
  encoder {IP} for {RISC-V} cores,''
  \url{https://github.com/accemic/TraceEncoder}, 2026, {CERN-OHL-S-2.0} OR
  Accemic commercial license.

\bibitem{shaheen}
L.~Valente \emph{et~al.}, ``A heterogeneous {RISC-V} based {SoC} for secure
  nano-{UAV} navigation,'' \emph{IEEE Transactions on Circuits and Systems I:
  Regular Papers}, 2024.

\bibitem{zgheib2022}
A.~Zgheib, O.~Potin, J.-B. Rigaud, and J.-M. Dutertre, ``A {CFI} verification
  system based on the {RISC-V} instruction trace encoder,'' in \emph{25th
  Euromicro Conference on Digital System Design (DSD)}.\hskip 1em plus 0.5em
  minus 0.4em\relax IEEE, 2022, {IEEE} Xplore document 9996921.

\bibitem{zgheib2023emfi}
------, ``Experimental {EMFI} detection on a {RISC-V} core using the {Trace}
  {Verifier} solution,'' \emph{Microprocessors and Microsystems}, vol. 103, p.
  104968, 2023.

\bibitem{gamino2020}
I.~Gamino~del R{\'i}o, A.~Mart{\'i}nez~Hell{\'i}n, {\'O}.~R. Polo,
  M.~Jim{\'e}nez~Arribas, P.~Parra, A.~da~Silva, J.~S{\'a}nchez, and
  S.~S{\'a}nchez, ``A {RISC-V} processor design for transparent tracing,''
  \emph{Electronics}, vol.~9, no.~11, p. 1873, 2020.

\bibitem{nexrv}
R.~Chyla and {RISC-V International, Nexus Trace Task Group}, ``{NexRv} ---
  {RISC-V} {N-Trace} reference encoder and decoder ({refcode/c}),''
  \url{https://github.com/riscv-non-isa/riscv-nexus-trace/tree/main/refcode/c},
  2024.

\bibitem{fzi-riscv-etrace}
{FZI Forschungszentrum Informatik}, ``{riscv-etrace} --- {Rust} {E-Trace}
  packet and tracer library,''
  \url{https://github.com/fzi-forschungszentrum-informatik/riscv-etrace}, 2025.

\bibitem{zak2025}
M.~Zak, M.~O'Donnell, and V.~Chickermane, ``Unleashing the power of {RISC-V}
  {E-Trace} with a highly configurable decoder,'' in \emph{RISC-V Summit
  Europe}.\hskip 1em plus 0.5em minus 0.4em\relax Paris, France: Siemens, 2025.

\bibitem{peakrdl}
A.~Nolting, ``{PeakRDL-regblock} --- {SystemRDL} to {SystemVerilog} register
  block compiler,'' \url{https://github.com/SystemRDL/PeakRDL-regblock}, 2026.

\bibitem{nexrv-ctte}
{Accemic Technologies GmbH}, ``{CEDARtools.TraceDecoder} --- {CTTD},''
  \url{https://github.com/accemic/CTTD/tree/main}, 2026.

\bibitem{sv2v}
Z.~Snow, ``{sv2v}: {SystemVerilog} to {Verilog} conversion,''
  \url{https://github.com/zachjs/sv2v}, 2026.

\bibitem{symbiyosys}
C.~Wolf and {YosysHQ}, ``{SymbiYosys} --- front-end for {Yosys}-based formal
  verification flows,'' \url{https://github.com/YosysHQ/sby}, 2026.

\bibitem{ntrace-refcompression}
{RISC-V International, Nexus Trace Task Group}, ``{N-Trace} compression tests
  --- reference-encoder results over the {E-Trace} test set
  ({\texttt{refcode/c/examples/all}}),''
  \url{https://github.com/riscv-non-isa/riscv-nexus-trace/tree/main/refcode/c/examples/all},
  2024, accessed 2026-08-11.

\bibitem{reuse}
{Free Software Foundation Europe}, ``{REUSE} {Specification} --- {SPDX}-based
  licensing and copyright compliance,'' \url{https://reuse.software}, 2024.

\bibitem{verilator}
W.~Snyder, ``{Verilator}: open-source {SystemVerilog} simulator and lint
  system,'' \url{https://verilator.org}, 2026.

\bibitem{abcflow}
{Accemic Technologies GmbH}, ``{abc-flow} --- project and build driver for
  {HDL} designs,'' \url{https://github.com/accemic/abc-flow}, 2026.

\end{thebibliography}

\end{document}